\documentclass{article}
\usepackage{epsfig} 
\usepackage{cite}  
\def\ln{\hbox{ln}}

\def\lapprox{\lower .7ex\hbox{$\;\stackrel{\textstyle <}{\sim}\;$}}
\def\gapprox{\lower .7ex\hbox{$\;\stackrel{\textstyle >}{\sim}\;$}}

\begin{document} 
\unitlength1cm 
\begin{titlepage} 
\vspace*{-1cm} 
\begin{flushright} 
ZU-TH 09/06\\
hep-ph/0604030\\
April 2006\\
\end{flushright} 
\vskip 2.5cm

\begin{center} 
{\Large\bf Measuring the Photon Fragmentation Function at HERA}
\vskip 1.cm 
{\large  A.~Gehrmann--De Ridder}$^{a}$, {\large  T.~Gehrmann}$^{b}$ 
and {\large E.~Poulsen}$^{b}$ 
\vskip .7cm 
{\it $^a$ Institute for Theoretical Physics, ETH, CH-8093 Z\"urich,
Switzerland} 
\vskip .4cm 
{\it $^b$ Institut f\"ur Theoretische Physik, Universit\"at Z\"urich,
Winterthurerstrasse 190,\\ CH-8057 Z\"urich, Switzerland} 
\end{center} 
\vskip 2.6cm 

\begin{abstract} 
The production of final state photons in deep inelastic scattering 
originates from photon radiation off leptons or quarks involved in the 
scattering process. Photon radiation off quarks involves a contribution 
from the quark-to-photon fragmentation function, corresponding to 
the non-perturbative transition of a hadronic jet into a single, highly 
energetic photon accompanied by some limited hadronic activity. Up to now,
this fragmentation function was measured only in electron-positron
annihilation at LEP. We demonstrate by a dedicated parton-level calculation 
that a competitive measurement 
of the quark-to-photon fragmentation function can be obtained in deep 
inelastic scattering 
at HERA. Such a measurement can be obtained  
by studying the photon energy spectra in 
$\gamma + (0+1)$-jet events, where
 $\gamma$ denotes a hadronic jet containing a highly energetic photon
(the photon jet). 
Isolated photons are  then defined from the photon jet by imposing a minimal 
photon energy fraction. For this so-called democratic clustering approach,
we study the cross sections for isolated 
$\gamma + (0+1)$-jet and  $\gamma + (1+1)$-jet production as well as 
for the inclusive isolated photon production in deep inelastic scattering.
\end{abstract} 
\vfill 

\end{titlepage} 

\newpage 

\renewcommand{\theequation}{\mbox{\arabic{section}.\arabic{equation}}}

\section{Introduction}
\setcounter{equation}{0}
The production of final state photons at large transverse momenta in high
energy processes provides an important testing ground for QCD.
A good understanding of the Standard Model predictions for 
photon production is essential for new physics searches at future
colliders. In high energy collisions, the produced 
primary partons, quarks or gluons, subsequently fragment 
into clusters of comoving hadrons, the hadronic jets. 
In events where a photon is produced in addition to the jets, this photon can
have two possible origins: the direct radiation of a photon off a primary
quark or antiquark (or, if leptons are also involved in the process, 
off a charged lepton)
and the fragmentation of a hadronic jet into a 
photon carrying a large fraction of the jet energy.
While the former direct process takes place  
at an early stage in the process of hadronisation and can be calculated 
in perturbative QCD, the
fragmentation contribution is primarily due to a long distance process 
which cannot be calculated within perturbative method. The latter is described
by the process-independent quark-, anti-quark- or gluon-to-photon 
fragmentation 
functions \cite{koller} which must be determined by experimental data.
Their evolution with the factorisation scale $\mu_{F,\gamma}$ can however be
calculated perturbatively. 
Furthermore, when the photon is radiated somewhat later during the
hadronisation process, in addition to this genuinely non-perturbative 
fragmentation process, the emission of a photon collinear to the primary 
quarks can also take place and has to be taken into account.
As physical cross sections are necessarily finite these collinear 
divergences will get factorised into the fragmentation functions. 
The factorisation procedure of these final state collinear singularities 
in fragmentation functions used here is 
of the same type as the procedure used to absorb initial state collinear 
singularities~\cite{AP} into the parton distribution functions.
 
Directly produced photons are usually well separated from the hadronic jets
produced in the event, while photons originating from the fragmentation
process and collinear quark-photon emission are primarily found 
inside hadronic jets.   
Consequently, it was thought that by imposing some isolation criterion 
one could eliminate the fragmentation process and define isolated photon
events in this way. However this is not the case: one can at most suppress 
the fragmentation and collinear contributions. In most theoretical 
observables involving final state photons, those contributions are 
indeed present.  

So far, only a limited number of measurements of single photon production
exists through which direct information on the quark-to-photon fragmentation
function (denoted by FF) can be obtained. A possible way is the measurement 
of inclusive photon cross sections in different experimental environments.
The OPAL Collaboration measured the inclusive photon rate \cite{opal} in 
$e^+e^-$ annihilation for $0.2<x_{\gamma}<1.0$ where in terms of the beam 
energy $x_{\gamma}=2E_{\gamma}/M_{Z}$ is the photon energy fraction.
The results were in reasonable agreement with predictions 
obtained using various model estimates of photon fragmentation functions 
for which the factorisation scale $\mu_{F,\gamma}$ was chosen to be equal 
to $M_{Z}$ \cite{duke,grv,bfg}.
The experimental precision was however 
not sufficiently high to discriminate between 
 different theoretical predictions.

An alternative way to determine the process-independent
photon fragmentation function is to measure the production of
photons accompanied by a definite number of hadronic jets.
It should be
noted that the quark-to-photon fragmentation function determined via
the measurement of inclusive or jet-like observables is the same in both
cases as it is process-independent. Indeed the fragmentation process
and the collinear quark-photon emission are found inside the hadron jets
and those contributions are the same whether one analyses
inclusive or jet-like  observables.

In processes involving hadronic jets and a photon in the final state, 
the outgoing photon is treated like any other hadron by the jet algorithm. 
It is clustered simultaneously with the other hadrons
into jets, within the so-called democratic procedure
\cite{glover-morgan,aleph}.
One of the jets will contain a photon and will be called photon-jet
if the fraction of energy carried by the photon $z$ inside the jet
is sufficiently large, i.e.
\begin{equation}
 z=\frac{E_{\gamma}}{E_{\gamma}+E_{had}} > z_{cut}\;,
\end{equation}
with $z_{cut}$ fixed experimentally.
$z$ can also be defined with respect to the transverse energies 
instead of the energies.

Following this line, the ALEPH Collaboration \cite{aleph} has analysed
events produced on  the $Z$-resonance in $e^+e^-$ collisions which contained
one hadron jet and one photon jet, where the photon carried
at least $70\%$ of the jet energy.
A comparison between the measured rate
and a leading order (LO), ${\cal O}(\alpha)$,
 calculation \cite{glover-morgan}
yielded a first determination of the quark-to-photon fragmentation function
in observables related to jets. It is worth noting that in this observable,
called the $\gamma +1$-jet rate, the quark-to-photon
fragmentation function appears already at the lowest order.
This observable is therefore highly sensitive on the quark-to-photon
fragmentation function and particularly suited to determine it.
The calculation of the $\gamma$+1-jet rate was furthermore extended
to next-to-leading order (NLO), i.e.\ up to (${\cal O}(\alpha \alpha_{s})$)
in \cite{agg} and a NLO fragmentation function
was obtained~\cite{ggg} by comparison with the ALEPH data.
Computing the inclusive photon rate in the same fixed-order framework 
with the LO and NLO fragmentation functions obtained from the ALEPH data,
one finds~\cite{gggopalaleph} that the 
results are in good agreement with the OPAL measurement~\cite{opal}.

To define isolated photons produced in a hadronic environment, 
a minimal amount of hadronic activity close to the photon must be admitted 
to ensure the infrared finiteness of the observable.
In the approach followed by ALEPH, 
the isolated photon rate 
is defined as the $\gamma$+1-jet rate where the photon carries 95 \% 
of the photon-jet energy. The amount of energy required for a photon 
inside the photon-jet to be called  ``isolated''  was fixed by 
analysing the data on the $\gamma$+1-jet rate for $0.7<z<1$.
The amount fixed depends on the experimental context.
%Different experimental environments may lead to an other amount of energy 
%required to define an isolated photon.  
In \cite{aleph}, the calculated and measured isolated rates 
were compared while varying the jet clustering parameter $y_{cut}$.
The theoretical prediction for the isolated rate defined as the 
$\gamma$+1-jet rate for $z>0.95$ using the measured photon 
fragmentation function at a given value of $y_{cut}$  were found 
in agreement with the measured isolated rate over 
the whole range of $y_{cut}$.
The inclusion of the NLO corrections in the theoretical prediction
improved the agreement.

Photon isolation from hadrons has been discussed intensively in the 
literature~\cite{frixione}, and  up to now 
the most common procedure  
uses a cone-based isolation criterion following the Snowmass convention 
\cite{snowmass}.
Recently, the ZEUS collaboration \cite{zeus} performed a measurement of
the inclusive isolated photon production cross section in electron-proton
deep inelastic scattering (DIS) 
at HERA using a cone-based isolation procedure. 
In this cross section, the photon carried $90\%$
of the energy inside a cone defined
in rapidity and azimuthal angle around the photon.
In \cite{zeus}, this measurement was compared to predictions
obtained with the Monte Carlo parton shower 
event generator programs PYTHIA~\cite{pythia}
and HERWIG \cite{herwig}, which do not include
photon fragmentation. ZEUS observed a noticeable excess
of the measurement compared to the predictions. Moreover, 
even after rescaling the normalisation of the cross section, none of the 
programs was able to describe all kinematical distributions 
of the experimental data in a satisfactory manner.

It was suggested in \cite{pisano}
that the isolated photon production cross section in 
DIS could be used to determine the photon distribution 
in the proton, 
assuming that all observed  isolated photons are radiated
from the lepton only. This photon distribution inside the proton 
is an important ingredient to 
electroweak corrections to cross sections at hadron 
colliders~\cite{phodist}.
 Although the observed total cross section 
seemed to be in agreement with model estimates based on QED-generated 
photon distributions in the proton~\cite{mrst}, it was recently 
demonstrated~\cite{saxon} that the kinematical distribution of photons
 inside the proton  can not be described in this approach. 

In \cite{zeusnew}, we performed a dedicated parton-level calculation
of the observable measured by ZEUS, using the same cone-based isolation
criterion as the ZEUS collaboration to define the isolated photon 
cross section. 
This parton-level calculation naturally includes two aspects which are 
neglected in the event generators: 
quark-to-photon fragmentation 
and large angle radiation of the photon from the lepton or from the quark.
Our results were found in good agreement with all aspects of the 
experimental measurement. 

In addition to measuring the inclusive isolated photon 
cross section,
the ZEUS collaboration also analysed~\cite{zeus} the production 
of prompt photons in association with hadronic jets. 
This measured cross section
was then compared with the NLO 
calculation \cite{herajet} of the $\gamma +(1+1)$-jet cross section: 
the cross section for the production of a photon-jet and one additional 
hadron-jet in the final state
($n$-jet observables in DIS are usually denoted by $(n+1)$-jet
  observables where the $+1$ stands for the unobserved jet coming from the 
proton remnant).
Data and theory were found in good agreement. 

For this observable however, the quark-to-photon fragmentation 
function enters only at the next-to-leading order. 
Indeed, the ZEUS Collaboration did not analyse their data in view of a
determination of the quark-to-photon fragmentation function but 
just compared data and theory for the $\gamma +(1+1)$-jet cross section. 

To measure the quark-to-photon fragmentation function at HERA in DIS, 
it seems best to
consider the analogue to the $\gamma +1$-jet rate at LEP, thus 
the $\gamma +(0+1)$-jet cross section. For this observable, besides 
the photon jet, no further hadronic jet activity is present in the final state
except the proton remnant jet, of course.
Moreover, the quark-to-photon fragmentation function enters at the 
lowest order.
It is the principal goal of this paper to advocate a measurement 
of the quark-to-photon fragmentation function utilising HERA data    
on $\gamma +(0+1)$-jet events in DIS.

More precisely, the plan of the paper is as  follows.
In section~\ref{sec:calc},
 we present the calculation of the 
 $\gamma +(0+1)$-jet cross section which consists of the hard photon emission
 and the fragmentation process and we discuss how these two contributions 
are combined and implemented 
into a parton-level Monte Carlo program.
Section~\ref{sec:frag} contains our predictions for the 
 $\gamma +(0+1)$-jet cross section 
differential in $z$  ($0.7 <z<1$) using a given jet algorithm to build 
$\gamma +(0+1)$-jet final states, evaluated for different
quark-to-photon fragmentation functions. We illustrate how a 
measurement of this differential 
cross section can be used to 
extract the  quark-to-photon fragmentation function.
Defining isolated photons in deep inelastic scattering by considering 
photon jets with $z>0.9$, in section~\ref{sec:isol},
we present our results for the 
isolated $\gamma +(0+1)$-jet cross section and 
the isolated inclusive photon cross section,  
differential in rapidity ($\eta_{\gamma}$)  and transverse energy 
($E_{T, \gamma}$). These are studied for different jet algorithms.
Finally, section~\ref{sec:conc} contains the conclusions and an outlook. 

\section{Parton-level calculation}
\setcounter{equation}{0}
\label{sec:calc}
We consider the production of $\gamma +(0+1)$ jets in DIS. 
$\gamma +(0+1)$ jets
are understood as a final state containing a highly energetic photon,
 which can be part of a hadronic jet (called the photon-jet and 
abbreviated by ``$\gamma$'' ), 
no further jet (``$+0$'') except the remnant jet (``$+1$'').
At leading order, the photon production process in DIS is  
${\cal O}(\alpha^{3})$ which is to be compared with ${\cal O}(\alpha^{2})$
for the inclusive deep-inelastic process. 
At this order, two different partonic processes yield $\gamma +(0+1)$-jet 
final states:
(a) $l q \to l q \gamma$, where the photon and the quark are either clustered 
together into a single jet ($z<1$) or the quark is well separated 
from the photon, but is at too low transverse momentum or at too large 
rapidity to be identified as a jet ($z=1$).
(b) $l q\to lq$ where the quark jet fragments into a highly energetic 
photon carrying a large fraction $z$ of the jet energy.
Both processes will be discussed in detail in  
the following subsections. Following those,
we  will describe how the two
contributions are combined and implemented in a numerical parton-level 
Monte Carlo program. 

\subsection{Kinematical definition of the observable}
To select  $\gamma +(0+1)$ jets in DIS, 
several criteria must be fulfilled by the 
final state particles:
deep inelastic scattering events (as opposed to 
photoproduction,~\cite{phoprodth,phoprodexp}) are 
selected by requiring the final state electron to be observed in the 
detector. The final state electron carries an energy $E_e$ and is observed 
at a scattering angle $\Theta_e$ (measured with respect to the incoming proton
direction). These variables  determine the common DIS variables 
$y$ and $Q^2$. The kinematics of the final state photon are characterised 
by its transverse energy $E_{T,\gamma}$ and its rapidity $\eta_\gamma$ 
(which may be inferred respectively from the transverse energy 
and the rapidity of the photon jet, defined by a jet algorithm). Finally, 
to avoid contributions from elastic Compton scattering $e p \to e p \gamma$, 
several hadronic tracks  are required in the detector. 

To define the $\gamma +(0+1)$ jet cross section in DIS, numerous cuts on the
kinematical variables for the final state electron and photon momenta are 
applied to preselect 
candidate events. In the following, we denote these cuts collectively by 
$\Theta(p_e,p_\gamma)$. The selected events are then subjected to a jet 
algorithm, 
which combines $n-1$ observed particle momenta, including the photon, 
and the proton remnant, (whose momentum
$p_n$ is inferred from momentum conservation),
into a 
$\gamma +(0+1)$-jet final state. We denote the action of this jet 
algorithm onto the $n$ final state momenta symbolically by a jet function
$J^{(n)}_{\gamma+(0+1)}(p_1,\ldots,p_n)$. 

\subsection{Hard photon emission processes}
\label{sec:a}
At leading order, ${\cal O}(\alpha^{3})$,  
the cross section for the production of hard photons  
in DIS
is described by the quark (antiquark) process 
\begin{displaymath}
l(p_1) + q(p_2) \rightarrow \gamma(p_3) + l(p_4) + q(p_5) 
\label{loprocess}
\end{displaymath}
with the particle momenta given in parentheses. $l$  denotes a lepton
or anti-lepton, and $q$ a quark or an anti-quark.
The momentum of the
incoming quark is a fraction $\xi$ of the proton momentum $P$, $p_2 =
\xi P$ and the proton remnant $r$ carries the momentum $p_r = (1 - \xi) P$. 
The latter hadronises into the remnant jet independently of the other 
final state particles.
The contribution of this process to the $\gamma +(0+1)$-jet cross section 
is given by
the integral over the three-parton final state phase space, weighted by 
the jet definition and the cuts:
\begin{equation}
\int {\rm d}PS_{3} \;|M|^2_{lq \rightarrow \gamma lq}
 \,J^{(3)}_{\gamma + (0+1)}(p_3,p_5,p_r) \, \Theta(p_3,p_4) \;.
\label{eq:m2three}
\end{equation}

\begin{figure}[t]
\begin{center}
\epsfig{file=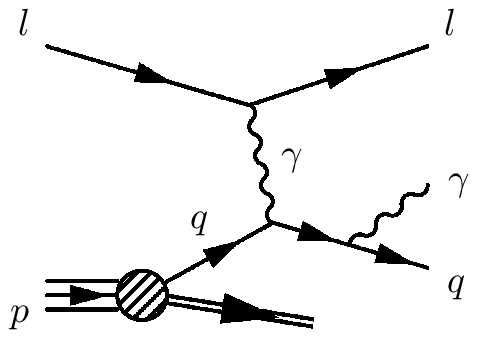,width=7cm}
\epsfig{file=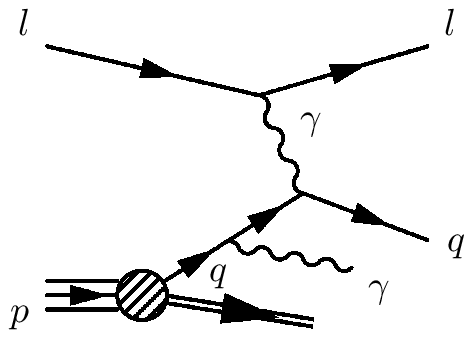,width=7cm}\\
\epsfig{file=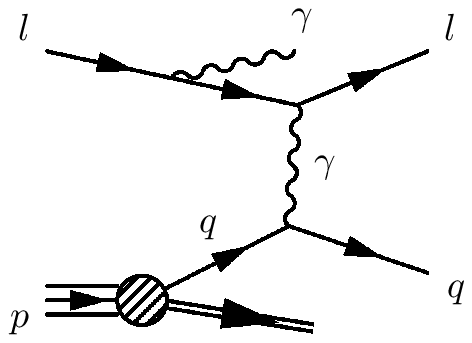,width=7cm}
\epsfig{file=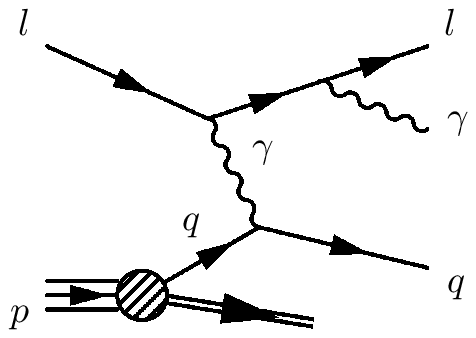,width=7cm}
\end{center}
\caption{Leading order Feynman amplitudes for hard photon production in
DIS. The $QQ$-contribution is obtained by squaring the 
sum of the upper two amplitudes, the $LL$-contribution 
from the square of the lower
two amplitudes, and $QL$-contribution from their interference.}
\label{fig:3p}
\end{figure}
Both leptons and quarks emit photons.
In the scattering amplitudes for this hard photon production process, depicted in Figure~\ref{fig:3p},
the lepton-quark interaction 
is mediated by the exchange of a virtual photon. The final state photon 
can be emitted off the lepton or off the quark. Consequently, one finds three
contributions to the cross section, coming from the squared amplitudes 
for radiation off the quark ($QQ$) or the lepton ($LL$), as well as the
interference of these amplitudes ($QL$). These contributions were computed 
originally as part
of the QED radiative corrections to DIS~\cite{riemann},
where the final state photon remains unobserved.  
The $QL$ contribution is odd under charge exchange, such that it contributes
with opposite sign to the cross sections with $l=e^-$ and $l=e^+$. 

In the $LL$ subprocess
the final state photon is radiated off the lepton. Since the 
cuts ensure that photon and electron are experimentally 
separated, this subprocess 
is free of a collinear electron-photon singularity. 
As the photon is radiated off the lepton, 
the momentum of the final state lepton can not be used to
determine the invariant four-momentum transfer between the lepton and 
the quark, which is in this subprocess given by 
$Q_{LL}^2= -(p_5-p_1)^2$, with $Q_{LL}^2 < Q^2 = Q^2_{QQ}
= -(p_4-p_2)^2$. 
In principle, $Q_{LL}^2$ is unconstrained by the kinematical cuts, and the 
squared matrix element for the $LL$
 subprocess contains an explicit $1/Q^2_{LL}$. 

In this process, the requirement of observing hadronic tracks comes into 
play, since the limit $Q^2_{LL}\to 0$ corresponds to photon radiation in 
elastic electron-proton scattering (also called Compton scattering),
$ep \to ep\gamma$. To translate the track requirement into parton-level 
variables, we proceed as discussed in~\cite{zeusnew}.  
The central tracking detectors of the HERA experiments cover 
in the forward region rapidities of $\eta < 2$. Requiring tracks in 
this region amounts to the current jet being at least partially contained 
in it. Assuming a current jet radius of one unit in rapidity, this amounts to 
a cut on the outgoing quark rapidity $\eta_q<3$, which we apply here.  
Varying this cut results only in small variations of the resulting cross 
sections. The cut on the 
outgoing quark rapidity enforces a minimum for $Q^2_{LL}$, thus it avoids a 
singularity in this subprocess cross section $\hat{\sigma}_{LL}$. 

Some care has to be taken 
in the choice of the factorisation scale for  the quark distribution 
function inside the proton, $\mu_F^2$, in the LL subprocess. 
In a leading order parton model
calculation, $\mu_F^2$ should ideally be taken to be 
the invariant four-momentum transfer to the quark, i.e.\ $Q_{LL}^2$ for the 
$LL$ subprocess. Even applying the quark rapidity cut, 
$Q^2_{LL}$ can assume low values, $Q^2_{LL}\sim \Lambda_{QCD}^2$, where the 
parton model description loses its meaning. Because of the cuts,
this kinematical region yields however only a small contribution 
to the cross section. 
To account for it in the parton model framework, we introduce a minimal 
factorisation scale $\mu_{F,{\rm min}} = 1$~GeV, and 
choose for the $LL$
subprocess $\mu_F$ = max($\mu_{F,{\rm min}}$,$Q_{LL}$), and for the 
$QL$ interference contribution $\mu_F$ = 
max($\mu_{F,{\rm min}}$,$(Q_{LL}+Q_{QQ})/2$).  
This fixed factorisation scale is 
an approximation to more elaborate procedures to extend the parton model 
to low virtualities~\cite{bk}, but sufficient in the present context.

This procedure for the scale setting in the 
$LL$ and $QL$ subprocesses is similar to what is done in the related
process of 
electroweak gauge boson production in electron-proton
collisions~\cite{bvz}. The major difference to~\cite{bvz} is that 
the cross section for isolated photon production in DIS vanishes for 
$Q^2_{QQ,LL} \to 0$, while being non-vanishing for vector boson production.
Consequently, in~\cite{bvz} the calculation of deep inelastic gauge 
boson production had to be supplemented by photoproduction of gauge 
bosons at $Q^2=0$, with a proper matching of both contributions at a
low scale. This is not necessary in our case. 

In the $QQ$ contribution, the photon radiated from the quark can have been 
radiated at two different stages of the hadronisation process.
The quark and the photon are usually well separated  
from each other if the radiation took place at an early stage, 
a process we shall name real hard emission.
When the photon is radiated somewhat later during the
hadronisation process, the emission of a photon collinear to the primary
quarks can take place which 
gives rise to a collinear singularity in the calculation.
Both contributions, hard and collinear emission processes, can be 
calculated within perturbative QCD as will be described below.

As physical cross sections are necessarily finite, the collinear
singularity appearing in the collinear emission process 
gets factorised into the fragmentation function defined at 
some factorisation scale $\mu_{F,\gamma}$.
The fragmentation process will be discussed in the next subsection.

In the real emission processes, the final state partons are 
experimentally unresolved, as quark and photon get clustered in one jet. 
Those partons  can be theoretically 
resolved, well separated from each other (real hard radiation)
or they can be theoretically 
unresolved. In the latter case the quark and the photon are collinear
(real collinear radiation).
The calculation of these two contributions is performed using the 
the phase space slicing method \cite{slicing}. 
By introducing a parameter $y_{min}$, one is able to separate 
the divergent, quark-photon collinear contribution from the 
finite contribution where the quark and the photon are theoretically separated.
 
The collinear contribution corresponds to 
the collinear limit of the matrix element integrated over the 
phase space region relevant to the collinear limit. This phase space region is
defined 
by $y_{q \gamma} <y_{min}$, where $y_{q\gamma} = s_{35}/s_{12}$ 
is the dimensionless invariant mass of the quark-photon system. 
Due to collinear factorisation of the phase space and the matrix element,
the collinear contribution yields a universal collinear factor 
multiplied by the 
hard $2\to 2$ cross section ($\hat{\sigma}_{e q \to e q}$).
This  divergent collinear factor is calculated analytically and absorbed into 
the quark-to photon fragmentation function as we will discuss in 
section \ref{sec:b}. Once this divergent part is factorised, 
the remaining two-parton process $e q \rightarrow e q$  
is evaluated numerically and  this collinear contribution 
yields always a $\gamma +(0+1)$-jet final state.
In obtaining the collinear factor, terms of order ${\cal O}(y_{min})$ 
have been neglected so that 
to obtain reliable results, $y_{min}$ is chosen to be small enough.
For our numerical results below, we shall use $y_{min}=10^{-7}$.

The finite contribution, where the quark and the photon are 
theoretically separated
is a three-parton process and  
is evaluated numerically for the three-parton phase space restricted by 
$y_{q\gamma} > y_{min}$. The jet algorithm is then applied to retain only 
$\gamma +(0+1)$ jet final states.
The $y_{min}$-dependence in the finite and collinear contributions 
cancels numerically when those are added together, 
such that the total $\gamma +(0+1)$-jet cross section is independent of this 
slicing parameter $y_{min}$. This independence yields an important 
check on the correctness of our calculation.

\subsection{Fragmentation contributions}
\label{sec:b}
In addition to the production of hard photons in the final state,
photons can also be produced through the fragmentation of a
hadronic quark jet into a single photon carrying a large fraction $z$ 
of the jet energy \cite{koller}. 
This fragmentation process is described in terms
of the quark-to-photon fragmentation function, $D_{q\to \gamma}$, 
which is convoluted with the cross section for the electron-quark 
scattering process
\begin{displaymath}
l(p_1) + q(p_2) \rightarrow l(p_4) + q(p_{35}) \;,
\end{displaymath}
such that the final state photon and quark momenta are given by 
$p_3 = z p_{35}$ and $p_5 = (1-z) p_{35}$. 

\begin{figure}[t]
\begin{center}
\epsfig{file=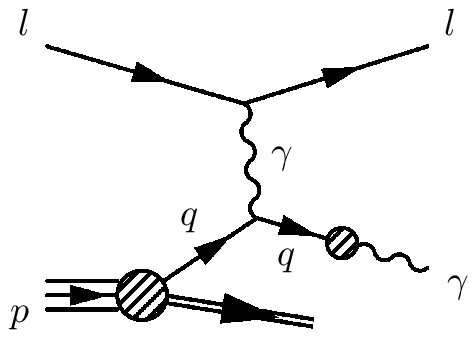,width=8cm}
\end{center}
\caption{Leading order Feynman amplitude for the quark-to-photon 
fragmentation process in deep inelastic scattering.}
\label{fig:2p}
\end{figure}

The fragmentation contribution to the $\gamma +(0+1)$-jet cross section 
associated with this fragmentation process
is displayed in Figure~\ref{fig:2p}. It  takes formally the following 
factorised form, 
\begin{equation}
\int {\rm d}PS_{2} \;|M|^2_{lq \rightarrow lq} 
 D_{q\to \gamma}(z) \,J^{(2)}_{\gamma+(0+1)} (p_{35},p_r)\, \Theta (p_3,p_4)\;.
\label{eq:m2frag}
\end{equation}
Here $J^{(2)}_{\gamma+(0+1)}$, 
the jet function defining how to obtain $\gamma+(0+1)$
jets out of one parton and the proton remnant, is simply $\Theta(z>z_{cut})$,
and thus independent of the jet recombination procedure. 

Like the hard photon contribution related to 
the parton process $l q \rightarrow l q \gamma$,
this fragmentation contribution is of order $\alpha^{3}$: 
The process $e q \rightarrow e q$ is of order $\alpha^2$ while the 
quark-to-photon fragmentation function $D_{q\to \gamma}$ is of 
order $\alpha$. The latter is given by,
\begin{equation}
D_{q \rightarrow \gamma}(z)= D_{q \to \gamma}(z,\mu_{F,\gamma}) + 
\frac{\alpha e_q^2}{2\pi}
 \left(P^{(0)}_{q\gamma}(z)\ln\frac{z(1-z)y_{min}s_{lq}}{\mu_{F,\gamma}^2} 
         + z\right)\, .
\label{eq:dqg}  
\end{equation} 
Here $D_{q\rightarrow \gamma}(z,\mu_{F,\gamma})$  stands for the
non-perturbative quark-to-photon fragmentation function describing 
the transition $q \rightarrow \gamma$ at
the factorisation scale $\mu_{F,\gamma}$. Parametrizations for this function
 will be specified below.
The second term in (\ref{eq:dqg}), if substituted in
(\ref{eq:m2frag}), represents the finite part obtained after absorption of 
the collinear quark-photon factor described in section~\ref{sec:a}
into the bare fragmentation function as explained in \cite{glover-morgan}.

In (\ref{eq:dqg}), $P_{q\gamma}^{(0)}$ is the LO quark-to-photon splitting
function
\begin{equation}
P_{q\gamma}^{(0)}(z) = \frac{1 + (1-z)^2}{z}
\label{pqgamma}
\end{equation}
and $e_q$ is the electric charge of the quark $q$. The variable $z$ denotes
the fraction of the quark energy carried away by the photon, while $s_{lq}$ 
is the lepton-quark squared centre-of-mass energy.

In order to turn the expression (\ref{eq:m2frag}) 
into a cross section,
one needs to know the non-perturbative quark-to-photon fragmentation function 
at the factorisation scale $\mu_{F,\gamma}$, 
$D_{q \to \gamma}(z,\mu_{F,\gamma})$.
This function satisfies an evolution equation 
which determines its variation with respect to the factorisation scale 
$\mu_{F,\gamma}$ to all orders in $\alpha_{s}$. 
Restricting ourselves to the zeroth order in $\alpha_{s}$,
at order $\alpha$, this fragmentation function obeys the leading order 
evolution equation, 
\begin{equation}
\frac{{\rm d}D_{q \to \gamma}(z,\mu_{F,\gamma})}{{\rm d}\ln \mu_{F,\gamma}^2}
=\frac{\alpha e_{q}^2}{2 \pi}P_{q\gamma}^{(0)}(z).
\label{evolution}
\end{equation} 
The fixed-order exact solution at ${\cal O}(\alpha)$ then reads,
\begin{equation}
 D_{q \to \gamma}(z,\mu_{F,\gamma})=\frac{\alpha e_{q}^2}{2 \pi} 
\,P_{q\gamma}^{(0)}(z)
\ln \left(\frac{\mu_{F,\gamma}^2}{\mu_{0}^2}\right) + D_{q \to \gamma}(z,\mu_{0})
\label{eq:npFF}
\end{equation}
$D_{q \to \gamma}(z,\mu_{0})$ is the quark-to-photon fragmentation function 
at some initial scale $\mu_{0}$. This function and the initial scale $\mu_0$ 
cannot be calculated and have to be determined from experimental data. 
First indications for a non-vanishing $D_{q \to \gamma}(z,\mu_{0})$
could be obtained by the EMC collaboration~\cite{emc} from the 
study of photon spectra in deep inelastic scattering, which were however 
insufficient for a detailed measurement. The first determination of 
$D_{q \to \gamma}(z,\mu_{0})$ was performed
by the ALEPH collaboration~\cite{aleph}. 
From their fit to the $e^+e^- \to \gamma +1$-jet 
data they obtained
\begin{equation}
D_{q \to \gamma}(z,\mu_{0})= \frac{\alpha e_{q}^2}{2 \pi} 
\left ( -P_{q\gamma}^{(0)}(z)\;\ln(1-z)^2 -13.26  \right ),
\label{eq:fit}
\end{equation}
with $\mu_{0}=0.14$ GeV. 
We note that (\ref{eq:npFF}) is an exact solution of (\ref{evolution}) 
at ${\cal O}(\alpha)$.
Furthermore when we substitute the solution (\ref{eq:npFF}) 
 into (\ref{eq:dqg}) the cross section 
becomes independent of the 
factorisation scale $\mu_{F,\gamma}$. 
This means that for the cancellation of the $\mu_{F,\gamma}$ dependence 
only the LO photon FF is needed.
Nonetheless, in order to see the influence of the NLO corrections to 
$D_{q \rightarrow \gamma}(z,\mu_{F,\gamma})$, we shall also evaluate 
the $\gamma+(0+1)$-jet 
cross section in DIS with the inclusion of the NLO photon FF. 

Similar to (\ref{eq:npFF}) the NLO fragmentation function 
 $D_{q \rightarrow \gamma}(z,\mu_{F,\gamma})$ is obtained as the solution 
of the evolution equation, but now with
O($\alpha \alpha_s$) terms added on the right-hand side of
(\ref{evolution}):
\begin{equation}
\frac{{\rm d}D_{q \to \gamma}(z,\mu_{F,\gamma})}{{\rm d}\ln \mu_{F,\gamma}^2}
=\frac{\alpha e_q^2}{2\pi}
\left[P^{(0)}_{q\gamma}(z)+
      \frac{\alpha_s}{2\pi} C_F P^{(1)}_{q\gamma}(z)\right]
+ \frac{\alpha_s}{2\pi} C_F P^{(0)}_{qq}(z)\otimes 
D_{q \rightarrow \gamma}(z,\mu_{F,\gamma}).
\end{equation}
The resulting quark-to-photon FF at scale $\mu_{F,\gamma}$ is 
\begin{equation}
\begin{array}{l} \displaystyle
D_{q \rightarrow \gamma}(z,\mu_{F,\gamma}) = \frac{\alpha e_q^2}{2\pi}
\left[P^{(0)}_{q\gamma}(z)+
      \frac{\alpha_s}{2\pi} C_F P^{(1)}_{q\gamma}(z)\right]
\ln\left(\frac{\mu_{F,\gamma}^2} {\mu_0^2}\right) 
\\ \displaystyle
 + \frac{\alpha_s}{2\pi} C_F P^{(0)}_{qq}(z)
\ln\left(\frac{\mu_{F,\gamma}^2}{\mu_0^2}\right) 
\otimes \left[\frac{\alpha e_q^2}{2\pi}
\frac{1}{2}P^{(0)}_{q\gamma}(z)
\ln \left(\frac{\mu_{F,\gamma}^2}{\mu_0^2}\right) +
D_{q \rightarrow \gamma}(z,\mu_0) \right] 
   + D_{q \rightarrow \gamma}(z,\mu_0).
\end{array}
\label{dqg-solu2}
\end{equation}
$P^{(1)}_{q\gamma}(z)$ is the next-to-leading order
quark-to-photon splitting function \cite{curci} and $P^{(0)}_{qq}(z)$ is
the  LO $qq$ splitting function~\cite{AP}. $D_{q \rightarrow
  \gamma}(z,\mu_0)$ is the initial value of the NLO FF, which contains
all unknown long-distance contributions. The result in (\ref{dqg-solu2})
is an exact solution of the evolution equation up to O($\alpha
\alpha_s$). The NLO photon FF has equally
been determined \cite{ggg} using the ALEPH $e^+e^- \to \gamma+1$-jet 
data\cite{aleph}. A three parameter fit with $\alpha_s(M_{Z}^2)=0.124$ 
yielded
\begin{equation}
D^{NLO}_{q \rightarrow \gamma}(z,\mu_{0})=\frac{\alpha e_q^2}{2\pi} \left(
-P^{(0)}_{q\gamma}(z) \ln(1-z)^2 + 20.8(1-z) - 11.07 \right)
\label{dqg-NLO}
\end{equation}
with $\mu_0 = 0.64$ GeV. Inside the experimental errors this fit for the
photon FF at $\mu_{0}$ describes~\cite{ggg} the ALEPH data at least as good as the LO fit
(\ref{eq:fit}). 

It should be noted that the above LO and NLO quark-to-photon 
FF do not take into account the resummation of
powers of $\ln (\mu_{F,\gamma}^2/\mu_0^2)$ as conventionally implemented, e.g.\ via the
Altarelli-Parisi evolution equations~\cite{AP}. Such resummations are only
unambiguous if the resummed logarithm is the only large logarithm in the
kinematical region under consideration. If logarithms of different
arguments can become simultaneously large, the resummation of one of
these logarithms at a given order implies that all other potentially
large logarithms are shifted into a higher order of the perturbative
expansion, i.e.\ are neglected. In the evaluation of the $\gamma+1$-jet
rate at ${\cal O}(\alpha)$ \cite{aleph} and ${\cal O}(\alpha \alpha_{s})$ 
\cite{agg} at
LEP for $0.7 < z < 1$, one encounters at least two different potentially
large logarithms, $\ln \mu^2_{F,\gamma}$  and $\ln (1-z)$. In the high-$z$
region, where the photon is isolated or almost isolated, it is
by far not clear that $\ln \mu^2_{F,\gamma}$ 
is the largest logarithm. Choosing 
not to resum the logarithms of $\ln \mu^2_{F,\gamma}$ is therefore equally
justified for the case of large $z$, $z \rightarrow 1$.

In the conventional approach, powers of $\ln (\mu_{F,\gamma}^2/\mu_0^2)$
are resummed. The parton-to-photon FF's $D_{i
 \rightarrow \gamma}(z,\mu_{F,\gamma})$ then satisfy a system of inhomogeneous
evolution equations \cite{AP}. 
The solution of these equations resums all leading logarithms of the type 
$\alpha_s^n \ln^{n+1}\mu_{F,\gamma}^2$. Including ${\cal O}(\alpha_{s})$ 
corrections to the splitting functions yields resummation of 
subleading logarithms of the type $\alpha_s^n \ln^n\mu_{F,\gamma}^2$. Several 
parametrisations of the photon FF are available in this approach.
These use some model assumptions to describe the 
initial FF at some low scale $\mu_0$.
The most recent parametrisation of the photon FF in this 
approach are the BFG 
fragmentation functions \cite{bfg}. This parametrisation has been compared 
to the ALEPH $\gamma$ +1-jet cross section which is sensitive
 to the large $z$ region ($0.7<z<1$) and found in agreement with the data
\cite{gggopalaleph}. 
Previous parametrisations were proposed in \cite{duke,grv}.
Those tend to predict the 
$\gamma +1$-jet cross section in excess compared 
to the ALEPH data and will not be considered in the remainder of this paper.

As already mentioned, the inclusion of the NLO quark-to-photon
fragmentation function in our evaluation of the $\gamma +(0+1)$-jet 
cross section is not required to cancel the 
factorisation scale dependence of the cross section. There,
only the leading order (LO) fragmentation function is required.
However,  
the NLO quark-to-photon fragmentation function corresponds 
to an expansion in $\alpha_{s}$ of the resummed quark-to-photon 
fragmentation function derived in a conventional approach 
\cite{gggopalaleph}, 
neglecting the initial fragmentation function  $D_{q \to \gamma}(z,\mu_{0})$.
It is therefore instructive to implement the 
NLO quark-to-photon fragmentation function
in the evaluation of the observable.   
Doing so will enable us to compare the results obtained 
in different approaches.

Thus we have three different quark-to-photon FF at our
disposal which have been compared and found in agreement with the ALEPH data: 
the fixed order LO 
parametrisation, using the ALEPH data directly to determine the
initial distribution given in (\ref{eq:npFF}) and (\ref{eq:fit}), a NLO  
determined function given by 
(\ref{dqg-solu2}) and (\ref{dqg-NLO}) directly fitted to the ALEPH data,
and the NLO parametrisation of BFG. 
A detailed comparison of the two approaches (fixed order and conventional) 
is given in \cite{gggopalaleph}. Results for the $\gamma +1$-jet rate for LEP 
as measured by ALEPH  with these different parametrizations of the photon FF 
are also shown in \cite{gggopalaleph}. 
Furthermore, results for the theoretical calculation of the 
$\gamma +(1+1)$-jet cross sections in DIS
using those various fragmentation functions are discussed in \cite{photonfrag}.

In the remainder of this paper, we shall use these three
parametrisations of the photon FF to predict differential $\gamma +(0+1)$-jet 
cross sections in DIS. Finally, for the numerical results presented in the 
following, we shall always use $\mu_{F,\gamma}^2 = Q^2$.

\subsection{Numerical implementation}
The  $\gamma +(0+1)$-jet cross section involves two partonic contributions:
the hard-photon production and quark-to-photon fragmentation processes. 
Consequently, the cross section which is evaluated numerically 
in the form of a parton-level Monte Carlo generator 
contains a three-parton channel and a two-parton channel. 
The three-parton channel 
is evaluated  with the restriction that the quark and the photon 
are theoretically resolved, i.e.\ not collinear, 
defined by  $y_{q \gamma} > y_{min}$. 
A recombination algorithm yielding a  $\gamma +(0+1)$-jet final state 
is then applied to the partons present in the final state  
and a $\gamma +(0+1)$-jet event  
is obtained as an event with a photon-jet and the proton remnant jet.
The two-parton channel is proportional 
to the quark-to-photon fragmentation function and contains the 
contribution from collinear quark-photon 
radiation, in the region $y_{q \gamma}\leq y_{min}$.
In this case, the final state partons always build a $\gamma +(0+1)$-jet event.
The partonic cross section for $\gamma +(0+1)$-jet production reads,
\begin{eqnarray}
\hat{\sigma} &=&\int_{y_{q \gamma}>y_{min}}{\rm d}PS_{3}\,
|M|^2_{l q \rightarrow lq \gamma}\;
J^{(3)}_{\gamma+(0+1)}(p_3,p_5,p_r) \, \Theta(p_3,p_4) 
 \nonumber \\
 &+ & \int {\rm d}PS_{2}\,
|M|^2_{l q \rightarrow lq}\; D_{q \to \gamma}(z)
\,J^{(2)}_{\gamma+(0+1)} (p_{35},p_r)\, \Theta (p_3,p_4)\;.
\label{sigma2}
\end{eqnarray}
%Here, $J_{(i)}^{(j)}$ is the jet-function which ensures that only 
%events with a $\gamma +(0+1)$-jet topology are taken into account to evaluate 
%the cross section. 
Contributions where the photon builds a jet on his own are 
also included in the first term of the above equation. These contributions 
are obtained if the quark is combined with the remnant or is at too low 
transverse momentum or at too large rapidity to be identified as a jet. 
The application of kinematical cuts on the outgoing electron and photon 
is formally given  
by $\Theta(p_{3},p_{4})$.
Details concerning the jet function and the kinematical cuts 
will be given in section \ref{sec:frag}. 

Finally, the cross section $\sigma$ for deep inelastic electron-proton 
scattering is obtained by a convolution between the parton-level 
cross section $\hat \sigma$ for a given quark flavour (\ref{sigma2}) 
with the corresponding parton distribution function
summed over all quark and anti-quark flavours.
For this, we use the the CTEQ6L \cite{cteq6l} leading order parametrisation 
of parton distributions.

\section{The $\gamma +(0+1)$-jet cross section}
\label{sec:frag}
\setcounter{equation}{0}
In this section, we present our predictions for the $\gamma +(0+1)$ jet
cross section in DIS at leading order,
i.e.\ to  ${\cal O}(\alpha^3)$. We focus in particular on the photon 
energy distribution of the photon jet by studying differential 
distributions in the photon energy fraction $z$. An experimental 
photon identification appears to be realistic only for large $z$: $0.7<z<1$. 
By comparing the predictions obtained with different 
parametrisations of the quark-to-photon 
fragmentation function, we will demonstrate the 
sensitivity of this observable on the photon FF.
From the measured differential cross section, these predictions 
could lead to a new determination of the quark-to-photon fragmentation 
function in DIS. 

We recall that a measurement of the 
photon FF in DIS from the $z$-distribution of the 
 $\gamma+(1+1)$-jet cross
section was suggested  in~\cite{photonfrag}. Compared to this, 
the measurement from the $\gamma+(0+1)$-jet cross section discussed here
has an important advantage. 
The photon fragmentation function enters here 
already at the leading order, while it enters as a higher-order correction 
to the $\gamma+(1+1)$-jet cross section. Consequently, the ratio of the 
$z>0.9$ contribution to the $0.7<z<0.9$ contributions is considerably 
larger in  $\gamma+(1+1)$-jet final states than in 
 $\gamma+(0+1)$-jet final states, which in turn renders the experimental 
separation of the different bins more difficult.

Before we present our results, we specify the kinematical 
selection criteria appropriate for the HERA experimental environment 
and give a brief description of the different jet algorithms used in our 
study.

\subsection{Kinematical selection criteria}  
\label{sec:kinfrag}
The results for the differential cross sections for the $\gamma +(0+1)$-jet 
cross section
 are obtained for energies and kinematical cuts appropriate for the 
HERA experiments \cite{carsten}. A combined data sample of incoming positrons 
and 
electrons is considered here, with a positron fraction of $85.6\%$.
The energies of the incoming electron
(or positron) and proton are $E_e = 27.5$ GeV and $E_p = 920$ GeV,
respectively. The cuts on the DIS variables are chosen as follows:
\begin{equation}
E_{e}> 10~{\rm GeV}\,, \quad  151^{\circ}< \Theta_{e}< 177^{\circ}\,,\quad
   Q^2 > 4~{\rm GeV}^2\, \quad \mbox{and} \quad y > 0.15   
\;. 
\label{cuts1}
\end{equation}
The cuts on the electron energy and the scattering angle are due to  experimental 
requirements for the unambiguous identification of the electron, 
reflecting the geometry of the H1 detector. The 
cut on $Q^2$ is intended to ensure deep inelastic scattering events, as 
opposed to photoproduction. As discussed earlier, this cut is effective on the
parton level only for the $QQ$ subprocess, while the $LL$ subprocess 
can involve much lower virtualities of the exchanged photon. 
Deep inelastic scattering kinematics in the $LL$ process are ensured 
experimentally by requiring multiple hadronic tracks in the final state, 
which we implemented by requiring a maximum rapidity of the outgoing quark 
$\eta_{q} <3$.
Finally, a 
cut on the energy transfer variable $y$ is part of the preselection of 
deep inelastic events, intended to minimise effects of 
electromagnetic radiative corrections. 
In our study, we choose the minimal value of $y$ considerably 
larger than in typical analyses in DIS: for the 
 $\gamma +(0+1)$-jet final states this large minimum value of $y$ enhances 
the importance of the fragmentation contribution relative to the hard 
photon radiation. 

Final states are classified as $\gamma +(0+1)$-jet events 
after a jet algorithm has been applied to the momenta of the final state 
hadrons and the photon. 
The photon is treated like the quark during the jet
formation according to 
the  so-called democratic procedure~\cite{glover-morgan}.
If a jet is formed,
it is called ``photon-jet'' if the photon carries a large fraction 
of the jet energy (or jet transverse energy) $z > z_{{\rm cut}}$.

For this observable, it is crucial to apply the 
jet algorithm in the HERA laboratory frame. This situation is 
different from most 
studies of $(n+1)$-jet production in DIS ($n\geq 2$), which  
are preferably performed in the $\gamma^*$-proton centre-of-mass frame. 
In these studies, the positive $z$-axis is chosen to be the proton direction,
proton and virtual photon are back-to-back and the produced hard jets
are also back-to-back in transverse momentum.
This is also the situation one faces when examining the production of    
$\gamma +(1+1)$ jets as described in \cite{photonfrag}. 
In this case, the transverse energy of the photon-jet is balanced 
against the transverse energy of the other hard jet in the final state.

However in the evaluation of the $\gamma +(0+1)$-jet production 
at leading order, there is no hard 
jet to be back-to-back to the photon-jet.
Indeed, if one views this observable in the 
$\gamma^*$-proton centre-of-mass frame, 
the quark-photon system  
is back-scattered in the negative $z$-direction. 
 Final state photon and quark are therefore
at vanishing transverse momentum, and a photon jet can not be defined in a 
sensible manner in this frame. 

In the HERA laboratory frame on the other hand, 
incoming proton and electron as well as the proton remnant move along the 
$z$-axis (with positive $z$-direction defined by the incoming proton).
The photon-jet has a transverse momentum  with respect to this axis,
which is counter-balanced by the transverse momentum of the outgoing 
electron. In this frame, jets are constructed using one of the jet 
algorithms explained below and described by their rapidity $\eta_j$ 
and transverse energy $E_{T,j}$ in the HERA frame. The rapidity of the photon 
jet $\eta_{\gamma-jet}$
is also called photon rapidity $\eta_\gamma$. 
One defines the photon energy fraction inside the photon-jet by
\begin{equation}
z=\frac{E_{T,\gamma}}{E_{T,\gamma-{\rm jet}}}\;. 
\end{equation}
On the level of the theoretical calculation an analoguous jet algorithm is applied   
to cluster the final state quark, photon and proton remnant 
into $\gamma +(0+1)$-jet final states. If photon and 
quark are clustered together to form the photon jet, we have the 
corresponding theoretical expression for the photon energy fraction 
inside the photon-quark cluster given by, 
\begin{equation}
z=\frac{E_{T,\gamma}}{E_{T,\gamma}+E_{T,q}}\;.
\end{equation}
While for photon and quark not being merged in the same jet, we always find 
$z=1$.

For our predictions we use $z_{cut}=0.7$ to identify a jet as photon jet. 
%The value of this cut is usually fixed in agreement with the experimental 
%conditions. 
Furthermore, cuts are imposed on the photon-jet itself.
The photon-jet is required to have a minimum transverse energy in the HERA 
frame, 
$E_{T,\gamma-jet}>3$~GeV and its rapidity is restricted to be 
$-1.2<\eta_{\gamma-jet}<1.8$. If photon and quark are not combined into a 
single jet, and the quark is also not combined with the proton remnant, 
we expect to have a $\gamma+(1+1)$-jet final state. However, this final 
state is observed only if the quark jet can be identified, i.e.\ has 
sufficient transverse energy $E_{T,q} > 2.5$~GeV and is inside the 
detector coverage ($-2.1 < \eta_q < 2.1$). If the quark is forming a 
jet on its own outside these quark jet cuts, one  still observes 
$\gamma+(0+1)$-jet final states. 

\subsection{Jet algorithms}

Concerning the jet formation itself, 
two kinds of jet algorithms are commonly 
used to study jet production in 
DIS~\cite{disjet1,disjet2}: 
the hadronic $k_{T}$-algorithm \cite{ellis-soper}, 
which was developed originally for 
hadron colliders, and a modified version of the 
Durham $k_{T}$-algorithm~\cite{durhamkt}, adapted for deep inelastic 
scattering~\cite{ktdis}. We briefly describe each algorithm in this section.

In the hadronic 
$k_{T}$-algorithm, which is applied here in the HERA laboratory frame, 
one computes for each particle $i$ and for each pair of particles 
$i,j$ the
quantities 
\begin{equation}
d_i = E_{T,i}^2\,, \quad d_{ij} = {\rm min} (E_{T,i}^2,E_{T,j}^2) 
\left((\eta_{i} -\eta_{j})^2 +(\phi_{i}-\phi_{j})^2\right)  /R^2 \;,
\end{equation}
where $\eta_i$ is the rapidity of particle $i$ and $\phi_i$ is its polar angle 
in the plane perpendicular to the incoming beam direction. $R$ is the
jet resolution parameter in this algorithm. One then searches the smallest of 
all $d_i$ and $d_{ij}$, which is labeled $d_{min}$. If $d_{min}$ is a 
$d_i$, then particle $i$ is identified as a jet and removed from the
clustering procedure. If $d_{min}$ is a $d_{ij}$, particles $i,j$ are merged 
into a new particle (proto-jet) with
\begin{equation}
E_{T,ij} = E_{T,i} + E_{T,j}\,,\quad
\eta_{ij} =  \frac{E_{T,i} \eta_i + E_{T,j}\eta_j}
{E_{T,ij}}\,,\quad
\phi_{ij} =  \frac{E_{T,i} \phi_i + E_{T,j}\phi_j}
{E_{T,ij}}\;.
\end{equation}
The algorithm is repeated until all remaining particles or proto-jets 
are identified as 
jets. Experimentally observable jets are then required to have some minimal 
amount of transverse energy $E_{T,min}$. All jets below $E_{T,min}$ are 
unobservable (and can thus be considered part of the proton remnant); the 
resolution parameter $R$ does therefore control how likely a low energy 
particle is clustered into the harder jets or into the remnant. 

Applied on the parton level, one computes 
\begin{equation}
d_{\gamma q}={\rm min}(E^2_{T,\gamma},E_{T,q}^2)\;
\left((\eta_{\gamma} -\eta_{q})^2 +(\phi_{\gamma}-\phi_{q})^2\right)  /R^2 
\end{equation}
and recombines photon and quark if 
\begin{equation}
d_{\gamma q} <  {\rm min}(E^2_{T,\gamma},E_{T,q}^2)\;.
\label{eq:ktcone}
\end{equation}
This condition can be expressed purely in terms of the angular distance of 
photon and quark:
\begin{equation}
(\eta_{\gamma} -\eta_{q})^2 +(\phi_{\gamma}-\phi_{q})^2  < R^2 \,.
\label{eq:cone}
\end{equation}
It should be noted that this simplified condition is valid only at the 
leading order, where the hadronic $k_T$-algorithm is applied only to two 
partons (quark and photon) and thus performs only a single iteration. As soon 
as more than two partons are present (at higher orders), the algorithm
iterates over all possible pairs of partons. It is noteworthy that 
(\ref{eq:cone}) is identical to the 
recombination condition  which is used in the cone algorithm~\cite{snowmass}
 in jet studies at hadron colliders and also in cone-based 
definitions of isolated photons. In these, the resolution parameter $R$ 
is the cone size. A detailed comparison of the hadronic $k_T$-algorithm 
and the cone algorithm can be found in~\cite{ellis-soper}.

If condition (\ref{eq:ktcone}) is fulfilled, quark and photon are 
recombined into a single photon jet at parton level, which has:
\begin{equation}
E_{T,\gamma-jet} = E_{T,\gamma} + E_{T,q}\,,\qquad
\eta_{\gamma-jet} =  \frac{E_{T,\gamma} \eta_\gamma + E_{T,q}\eta_q}
{E_{T,\gamma-jet}}\;.
\end{equation}

The modified Durham $k_{T}$-algorithm~\cite{ktdis}, 
also applied in the HERA laboratory frame and  
adapted to the application in DIS features an important difference to the 
original formulation for $e^+e^-$ annihilation: 
the proton remnant is taken into account in the jet 
formation. For this algorithm, we 
consider the exclusive and inclusive 
formulation: the inclusive $k_{T}$-algorithm clusters until only the 
desired $\gamma+(0+1)$-jet final state is left, while the 
exclusive $k_{T}$-algorithm stops the recombination of particles 
according to a jet resolution parameter. 
A detailed discussion of both options for jet production in DIS can
be found in~\cite{disjet1,disjet2}.

Both inclusive and exclusive $k_T$-algorithms applied in the HERA laboratory
 frame 
calculate the quantity 
\begin{equation}
E^2_{T,ij}=2\,{\rm min}(E^2_{T,i},E^2_{T,j})(1-\cos \theta_{ij})
\label{eq:ktdur}
\end{equation}
for each pair $i,j$  of particles. The pair with the lowest 
$E^2_{T,ij}$ is then combined into a new particle by adding the momenta 
of $i$ and $j$. For the inclusive $k_{T}$-algorithm, this procedure is 
repeated until only a $\gamma+(0+1)$~jet final state is left, while for 
the exclusive $k_{T}$-algorithm, the procedure stops as soon as the 
pair with the lowest $E^2_{T,ij}$ has $E^2_{T,ij}/W^2 <y_{cut}$, where 
$W^2$ is the total invariant mass of the hadronic final state including the 
photon.  $y_{cut}$ is the experimental jet resolution 
parameter, it determines the broadness of the jet. It has in fact a similar 
role as $R$, the resolution parameter of the hadronic $k_T$-algorithm, or 
the radius of the cone in the cone algorithm. 

On the parton level, we compute (\ref{eq:ktdur}) for $i,j$ 
being each pair of two of the three partons: photon, quark and proton remnant.
The jet algorithm then selects the minimum of these three quantities.
In the inclusive case, the pair $i,j$ of partons with the minimal value of 
$E^2_{T,ij}$ is always combined, while for the exclusive case
this pair is combined only if 
$E^2_{T,ij}/W^2 <y_{cut}$ 
with $W^2$ being the squared invariant  mass of photon, quark and proton
 remnant.

Quark and photon build one jet if the minimal value of  $E^2_{T,ij}$ 
is given by 
$E^2_{T,\gamma q}$ (in the exclusive case, $E^2_{T,\gamma q}
/W^2 <y_{cut}$ has to be fulfilled as well). 
If photon and remnant are combined, the event is 
always discarded, while it is always accepted if quark and remnant 
are combined. In this case, the photon forms a jet on its own. In the 
exclusive case, we can have photon, quark and remnant forming each a 
jet on their own, i.e.\ yielding a $\gamma +(1+1)$-jet final state.
As we will see, for large values of $y_{cut}$ (above $y_{cut}=0.1$) 
both inclusive and exclusive jet algorithms lead to very similar predictions 
for the $\gamma +(0+1)$-jet cross section.

If photon and quark are combined, we compute for the photon jet
\begin{equation}
E_{T,\gamma-jet} = E_{T,\gamma} + E_{T,q}\,,\qquad
\eta_{\gamma-jet} = \frac{1}{2} \log 
\frac{E_\gamma + E_q + p_{z,\gamma}+ p_{z,q}}
{E_\gamma + E_q - p_{z,\gamma}- p_{z,q}} \;.
\end{equation}

We finally 
recall that applying any of the 
jet  algorithms on the parton level will classify two types of 
partonic contributions 
as  $\gamma +(0+1)$-jet final states. Events where the 
quark and photon are recombined into a 
jet will have $z<1$. 
On the other hand events where the photon forms a jet on its own 
while the quark is combined with the remnant or is produced at 
too low transverse energy or too large rapidity to be observed as 
a jet are also identified as $\gamma +(0+1)$-jet, with $z=1$.

\subsection{Measuring the quark-to-photon fragmentation function}
\begin{figure}[t]
\begin{center}
\epsfig{file=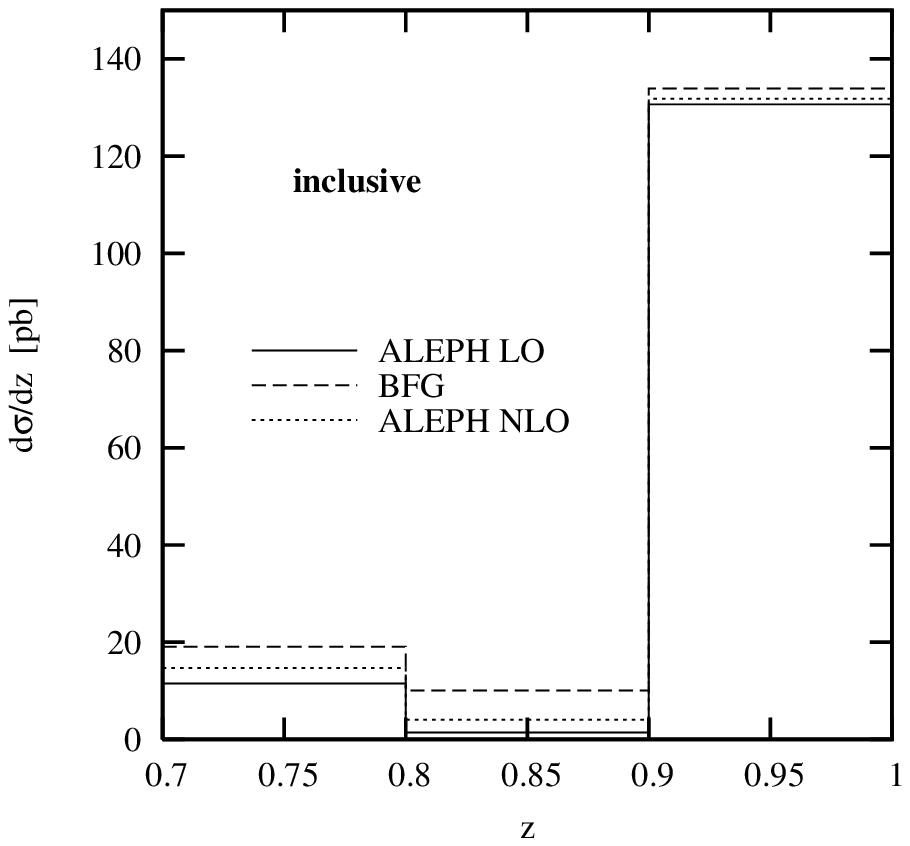,angle=-90,width=8cm} 
\hspace{-1.5cm} \epsfig{file=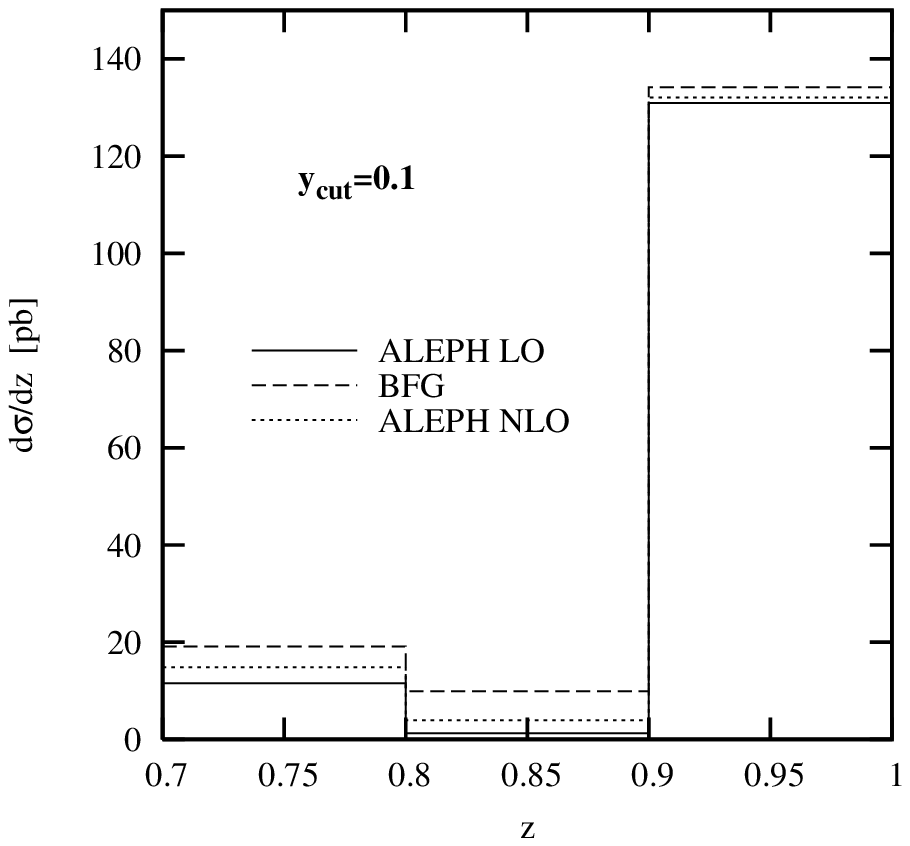,angle=-90,width=8cm}
\\[1cm]
\epsfig{file=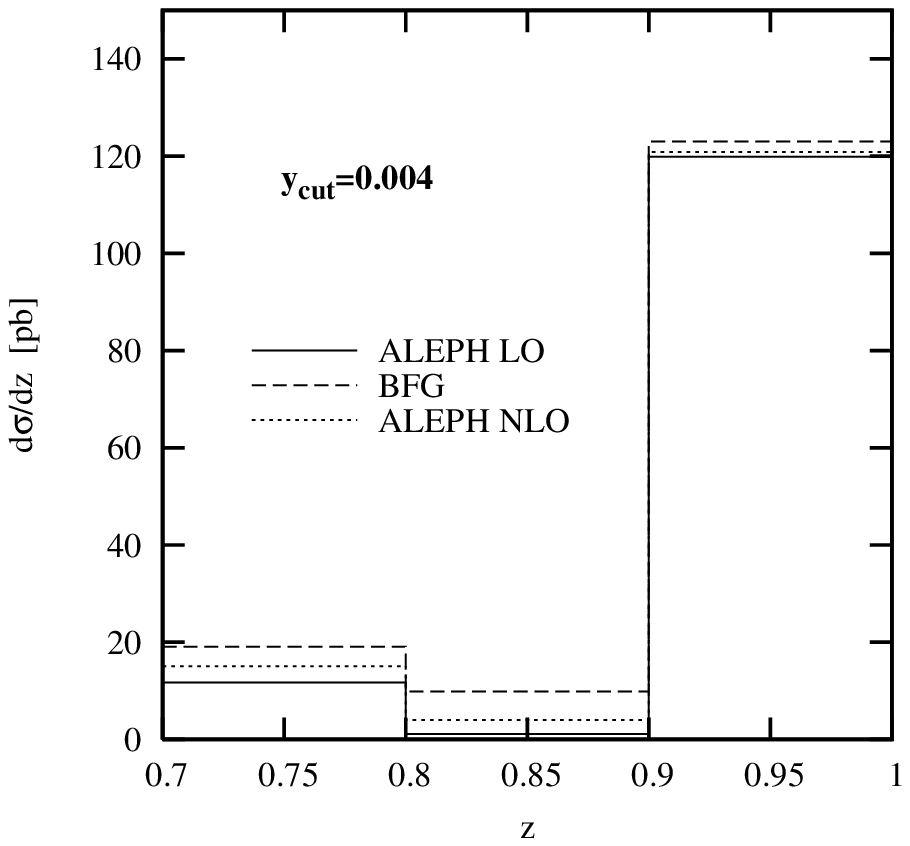,angle=-90,width=8cm} 
\hspace{-1.5cm} \epsfig{file=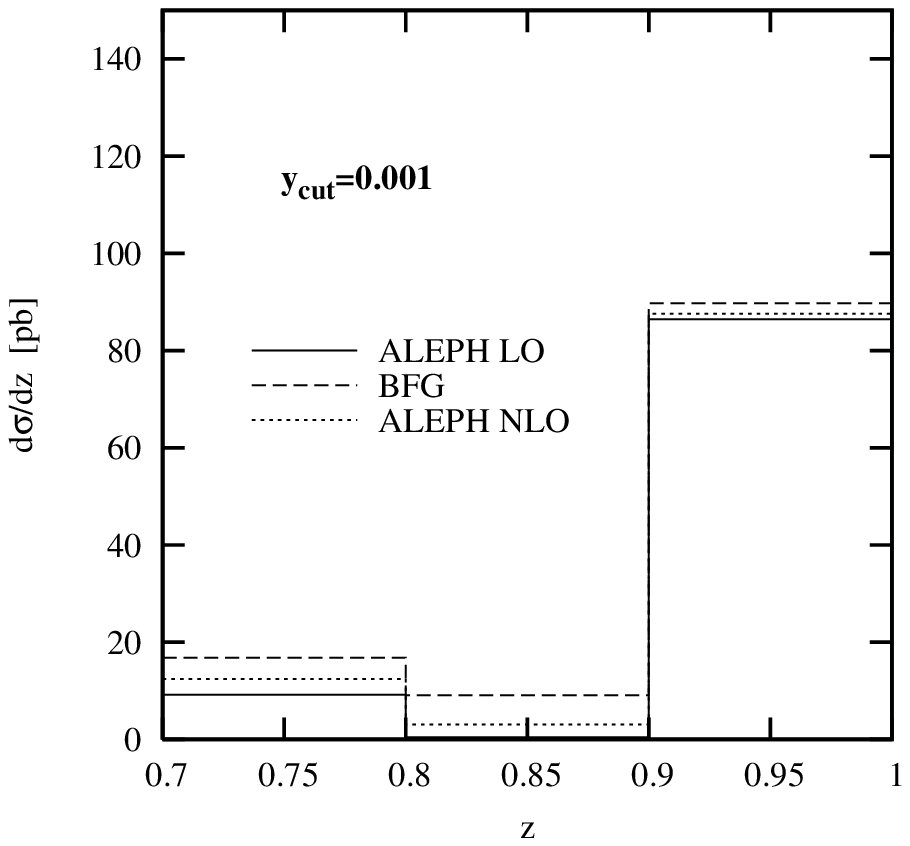,angle=-90,width=8cm}
\end{center}
\caption{Photon energy distribution inside the photon jet 
of $\gamma + (0+1)$-jet events. Jets are defined using the inclusive 
and exclusive $k_T$-algorithm. In the latter case the jet resolution parmeter 
$y_{cut}$ is taken equal to 0.1,0.004 and 0.001 respectively}
\label{fig:z}
\end{figure}

Predictions for the $\gamma +(0+1)$-jet cross section differential in $z$, 
obtained using the kinematical cuts specified in 
section~\ref{sec:kinfrag} and defined using the jet algorithms  
in the laboratory frame, are displayed in 
Figure~\ref{fig:z}. We use the three different parametrisations of the
photon fragmentation functions discussed in section~\ref{sec:b}
and apply either the inclusive or exclusive 
$k_T$-algorithm for different values of the resolution parameter 
$y_{cut}$. Results are given as bin-integrated cross sections for three 
bins, as anticipated~\cite{carsten} for the experimental 
measurement.

Concerning the variation with the jet resolution parameter, we 
observe that the inclusive $k_T$-algorithm and the exclusive $k_T$-algorithm
for large $y_{cut} = 0.1$ and above 
yield very similar results, indicating that for 
large $y_{cut}$, practically all events are classified as 
$\gamma +(0+1)$ jet. A visible variation of the cross section is 
observed only at much lower $y_{cut}$, $y_{cut}\ll 0.01$. 
For $y_{cut} = 0.004$ and even more for $y_{cut} = 0.001$, 
we observe that the $z>0.9$ contribution 
decreases considerably, while the contributions for lower $z$ remain 
largely unmodified. This can be understood from the fact that with 
decreasing $y_{cut}$, particles are less likely to be recombined into jets. 
In our case, especially quark and remnant are combined less often, 
such that more events at $z=1$
are classified as $\gamma +(1+1)$-jet events, resulting in a decrease of the 
$\gamma +(0+1)$-jet cross section in the last bin in $z$.

If we compare our results for the various photon FFs in Fig.\
\ref{fig:z},  we observe that the predictions
agree approximately within $5\%$ in the large $z$
region, i.e.\ for $z>0.9$. 
However,
near the minimum of the cross section, i.e.\ in the region
$0.7<z<0.9$, the results differ considerably by up to a factor 
2 in $0.7<z<0.8$ and up to a factor 5 in $0.8<z<0.9$. The
largest differences occur between the predictions obtained with the
LO ALEPH photon fragmentation function on the one hand and the BFG
parametrisation on the other hand. This discrepancy comes mainly from
the fact that different evolution approaches are used. Whereas for 
BFG the FF at $\mu_{F,\gamma}^2=Q^2$ is obtained from the conventional
evolution resumming the leading and subleading logarithms of $\mu_{F,\gamma}$, 
the ALEPH photon FFs are evolved only to the respective finite
order in $\alpha_{s}$ as given in (\ref{eq:npFF}) and
(\ref{dqg-solu2}). Therefore, if we calculated  
the $\gamma +(0+1)$-jet cross section 
at the large scale $\mu_{F,\gamma}^2=M_Z^2$ 
the cross sections obtained for BFG and ALEPH would come out quite
similar over the whole $z$-range inside a $20\%$ margin. Only
when we go to the scale $\mu_{F,\gamma}^2=Q^2$, which is much
smaller than $M_Z^2$, 
we observe that the cross section obtained using the BFG photon
fragmentation function is much larger than the ALEPH cross section in
the region $0.7 < z <0.9$. This was already observed 
in \cite{photonfrag} when comparing the predictions obtained for the 
$\gamma + (1+1)$-jet cross section in DIS.

Since the non-perturbative input distributions and 
higher order splitting functions contain explicit $\log(1-z)$ terms, 
it is however not clear if the resummed fragmentation functions can be 
considered to be reliable for $z>0.95$~\cite{gggopalaleph}. 
Provided the resummed solution of the evolution equation is 
accurate over the whole $z$ 
range under
consideration, i.e.\ for $0.7<z_{\gamma}<1$,
the approach using this
solution represents the theoretically preferred one as it is the
most complete. The fixed order
 approach using an expanded and therefore approximated
photon FF has on the other hand 
also important advantages. As already mentioned, its
use leads to factorisation scale independent results for the cross
section evaluated at a given fixed order in $\alpha_{s}$. Moreover it
enables an analytic determination of the photon FF. 

As the predictions for the $\gamma+(0+1)$-jet cross section
obtained using different parametrisations differ considerably, this observable 
is highly sensitive on the photon FF and would be an appropriate 
observable to measure in view of extracting
the quark-to-photon fragmentation function in DIS. 
Such a measurement could also be used to test the existing approaches to the 
FF discussed in section~\ref{sec:b}.

Finally, using the hadronic $k_T$-algorithm, the fraction of 
events where photon and quark are clustered together is considerably 
smaller than the one obtained using the exlusive (or inclusive)  
$k_T$-algorithm in the laboratory frame. As a 
consequence, more events are in the last bin $z>0.9$, and the fraction of 
events in the two other bins becomes negligible. Using either 
fragmentation function, one observes  that,
for the hadronic $k_T$-algorithm with 
separation parameters $R\leq 1$, negative contributions are predicted for 
the bins $0.7<z<0.8$ and $0.8<z<0.9$ for all fragmentation functions 
considered in this paper. These unphysical predictions 
can be understood as follows: one 
observes two types of logarithms in the 
$z$-distribution of the $\gamma +(0+1)$-jet cross section:
 $\ln(E^2_{T,\gamma}/Q^2)$ 
and $\ln(k^2_{T,\gamma-q}/Q^2)$, where $k_{T,\gamma-q}$ is the 
maximum
transverse momentum of the quark and the photon with respect 
to the photon jet direction allowed by the jet algorithm. 
While the former logarithms do not become 
large, since $E^2_{T,\gamma}$ and $Q^2$ are typically of the same magnitude, 
the latter logarithms can become large, if the jet algorithm is 
too restrictive in recombining quark and photon. In the case of the 
hadronic $k_T$-algorithm, $k^2_{T,\gamma-q}$ becomes much smaller 
than typical hadronisation scales for large $z$, and either approach 
(fixed order or resummed) to the photon fragmentation function loses 
its applicability. For the ALEPH NLO and BFG parametrisations, this effect 
may be accounted for in part by lowering the factorisation scale 
$\mu_{F,\gamma}$ associated with the photon fragmentation process,
but for large $z$, $k^2_{T,\gamma-q}$ in the hadronic $k_T$-algorithm
is too low to be taken as $\mu_{F,\gamma}$. Therefore, a measurement 
of the photon fragmentation function from the
$\gamma+(0+1)$-jet cross section should be based on the HERA-frame 
exclusive (or inclusive) $k_T$-algorithm, 
which admits larger values of $k^2_{T,\gamma-q}$, thus 
avoiding the appearance of the above-mentioned large logarithmic corrections.

\section{The isolated $\gamma+(0+1)$-jet and inclusive $\gamma$  
cross sections} 
\label{sec:isol}
\setcounter{equation}{0}

Production of isolated photons in association with hadrons
has been widely studied in different 
collider environments. The measured isolated photon cross sections were 
used as tests of the hard interaction dynamics, or to measure 
auxiliary quantities such as parton distributions. 
 A very sensitive issue is the 
definition of isolated photons produced in association with hadrons, since 
a completely isolated photon is not an infrared safe observable in 
quantum chromodynamics (QCD). At present, this isolation is 
usually accomplished experimentally 
by admitting a limited amount of hadronic energy inside a cone 
around the photon direction. 

The ZEUS study of isolated photon production in deep inelastic 
scattering~\cite{zeus} 
was carried out using such 
a cone-based isolation 
criterion, requiring the photon to carry at least 90\% of the 
energy inside a cone of unit radius in rapidity and polar angle, thus 
admitting 10\% of hadronic energy. We showed in~\cite{zeusnew} that the 
ZEUS measurement could be well reproduced in all its aspects
by a parton-level calculation, 
closely related to the calculation of the 
$\gamma +(0+1)$-jet cross section described above. In fact, the isolated 
photon cross section can be obtained from (\ref{sigma2}) by replacing
the $n$-particle jet functions $J^{(n)}_{\gamma+(0+1)}(p_3,p_5,p_r)$ 
by a photon isolation definition $I^{(n)}_{\gamma}(p_3,p_5)$. The 
cone-based isolation definition $I^{(3)}_{\gamma}(p_3,p_5)$
checks if the quark momentum $p_5$ 
is inside the cone defined by the photon momentum $p_3$, 
and subsequently applies a cut on the photon energy fraction $z>z_{cut}$. 
Since in the two-parton contribution, quark and photon momenta are 
always collinear, $I^{(2)}_{\gamma}(p_3,p_5)$ amounts simply to a cut 
on the photon energy fraction $z>z_{cut}$. The cross section for isolated 
photon production is thus also dependent on the photon fragmentation function.
As demonstrated in~\cite{zeusnew}, its prediction is however only marginally 
sensitive on the parametrisation used for the photon fragmentation function. 
In the following, we will therefore compute all predictions using just the 
ALEPH LO fragmentation function.

The cone-based isolation criterion has several conceptual drawbacks. The cone 
size can not be chosen much smaller than unity~\cite{Catani-Fontannaz}, 
as often required for new particle searches, since a small cone size would 
spoil the convergence of the perturbative expansion for the isolated photon 
cross section. The interplay of the isolation cone with other kinematical 
cuts can also sometimes lead to a discontinuous behaviour of the cross
section~\cite{binoth}.
Also, when studying the production of photons in 
association with hadronic jets, the application of the cone-based 
photon isolation could become ambiguous, since is is not 
clear how to attribute the hadronic activity in the photon isolation cone 
to the jets. 

To circumvent the problems of the cone-based photon isolation,
several alternative photon isolation criteria 
were proposed in the literature. A dynamic cone-based isolation~\cite{frixione}
could in principle allow to eliminate the dependence on the photon 
fragmentation function; this was however not accomplished in an experimental 
measurement up to now. In the democratic clustering procedure proposed 
in~\cite{glover-morgan}, isolated photon cross sections are 
directly derived from jet cross sections. In this approach, which was 
already used to define the $\gamma +(0+1)$-jet cross section in 
section~\ref{sec:frag},
the jet algorithm treats the photon like any other hadron, resulting 
it to be clustered into one of the final state jets, which is then
called the photon jet. An isolated photon in this approach is a 
photon jet where the photon carries more than a certain fraction of the 
jet energy. Using this democratic clustering approach, the ALEPH 
collaboration measured the isolated $\gamma+1$-jet rate~\cite{aleph} for the 
$k_T$-algorithm~\cite{durhamkt},
using $z_{cut}=0.95$ to define isolated photons. 
Using the fragmentation function previously determined from the photon energy
spectra of the $\gamma+1$-jet rate~\cite{aleph}, good agreement between 
experimental data and theory was found for a wide range of jet resolution 
parameters. This agreement improved considerably by including NLO 
corrections~\cite{agg,ggg}.

In this section, we study isolated photon cross sections in DIS, 
obtained using different jet algorithms. In contrast to the 
discussion of the previous section, where we aimed to maximise the sensitivity 
of our observable on the photon fragmentation function by restrictive cuts and 
by choosing a specific jet algorithm, here we choose a less restraint 
event selection. As before, we assume a combined data sample 
of incoming positrons and 
electrons, with a positron fraction of $85.6\%$, with 
 $E_e = 27.5$ GeV and $E_p = 920$ GeV.
Our choice of cuts is again motivated by the 
coverage of the H1 detector~\cite{carsten}.
In particular, we apply the following cuts 
on the DIS variables
\begin{equation}
E_{e}> 10~{\rm GeV}\,, \quad  151^{\circ}< \Theta_{e}< 177^{\circ}\,,\quad
   Q^2 > 4~{\rm GeV}^2\, \quad \mbox{and} \quad y > 0.05   
\;. 
\label{cuts2}
\end{equation}
Events selected using these criteria and containing a photon candidate 
are then processed using a jet algorithm. We have seen in the previous 
section that the difference between the inclusive and 
exclusive laboratory frame $k_T$-algorithm is only marginal, except for 
very small jet resolution parameters $y_{cut}$. For our studies here, we 
do therefore use only the exclusive laboratory frame $k_T$-algorithm
with $y_{cut} = 0.1$
and the hadronic $k_T$-algorithm with jet resolution parameter 
$R=1$. Both jet algorithms result in final states
containing a number of hard jets, with one of the jets containing 
the photon candidate. If the photon carries more than 90\% 
of the transverse energy of this photon jet ($z_{cut}=0.9$), 
it is called isolated. 
We then apply cuts on the photon transverse energy $E_{T,\gamma}$ and 
the photon rapidity $\eta_\gamma$:
\begin{equation}
E_{T,\gamma} > 3~\mbox{GeV}\;, \qquad -1.2< \eta_\gamma < 1.8\,.
\end{equation}
\begin{figure}[t]
\begin{center}
\epsfig{file=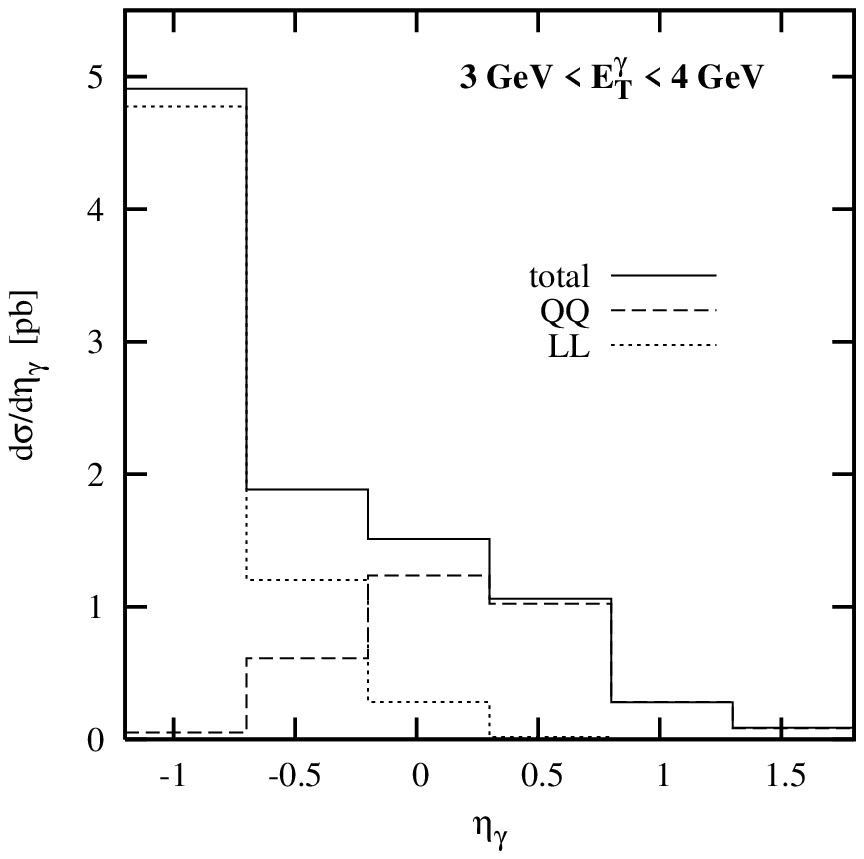,angle=-90,width=8cm} 
\hspace{-1.5cm} \epsfig{file=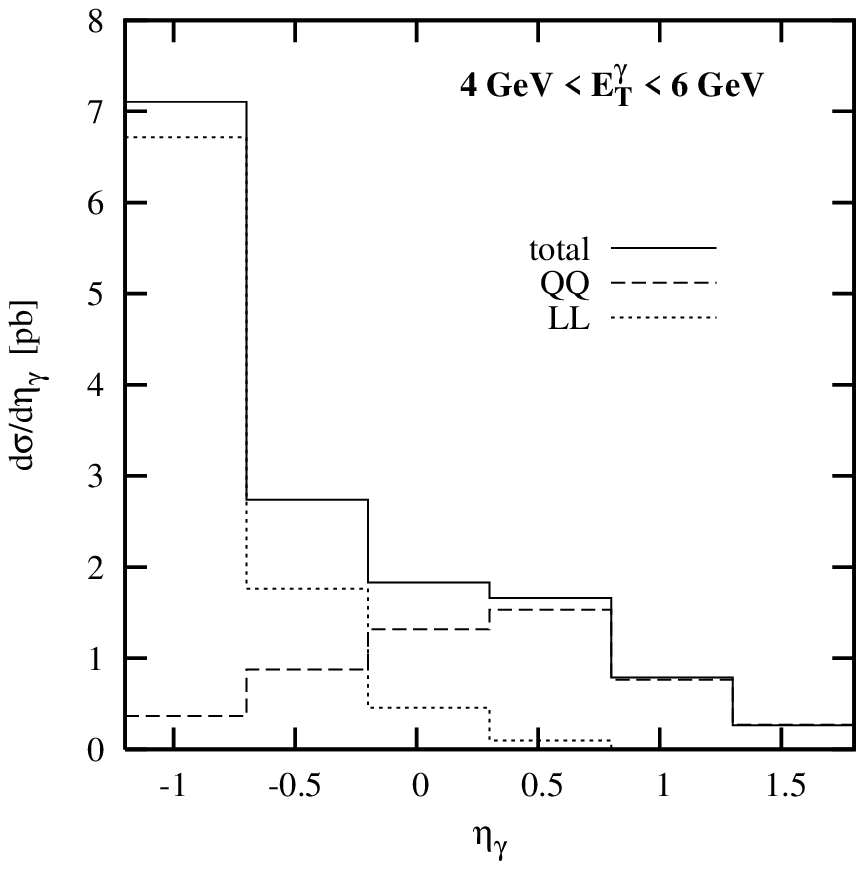,angle=-90,width=8cm}
\\[1cm]
\epsfig{file=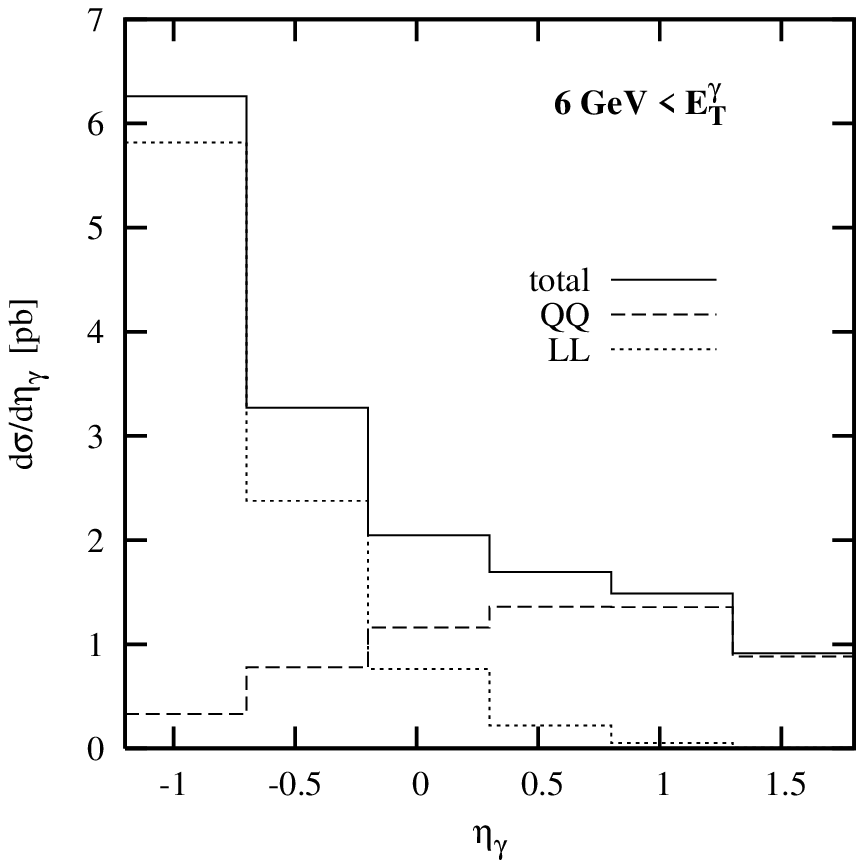,angle=-90,width=8cm} 
\hspace{-1.5cm} \epsfig{file=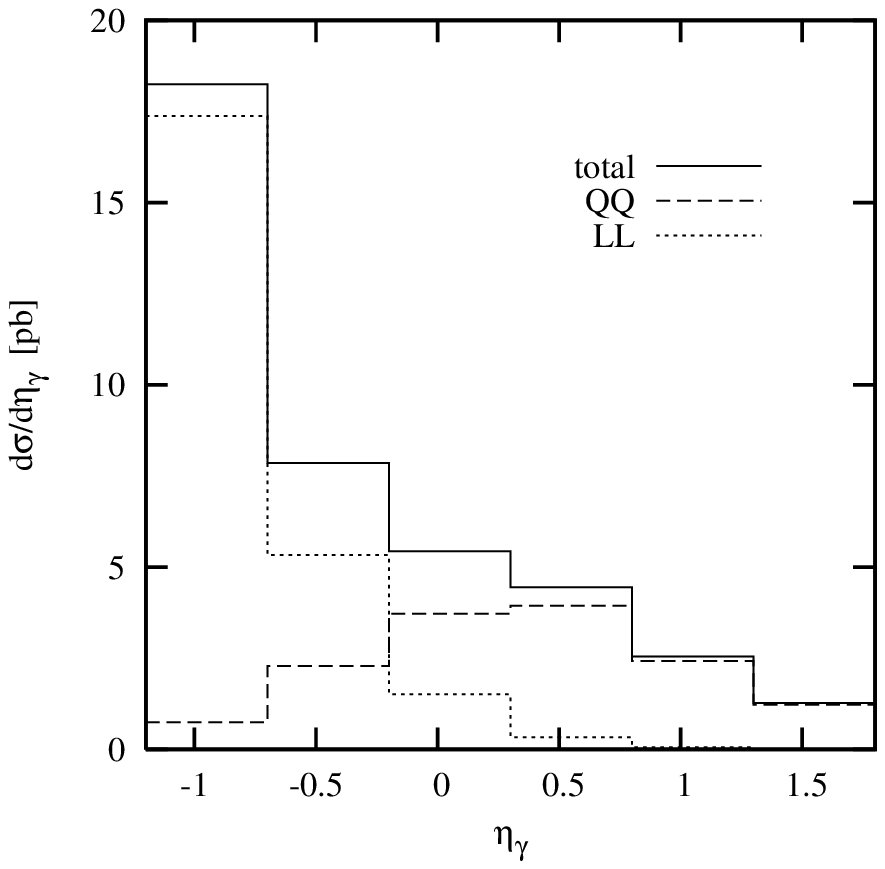,angle=-90,width=8cm}
\end{center}
\caption{Rapidity distributions of isolated photons in $\gamma+(0+1)$-jet 
events, in different bins 
in $E_{T,\gamma}$. The last plot shows the sum over 
all bins. Isolated photons are 
defined here 
using the exclusive $k_T$-algorithm ($y_{cut} = 0.1$) in the HERA frame, 
requiring $z>0.9$. $LL$ and 
$QQ$ subprocess contributions are indicated as dashed and dotted lines.}
\label{fig:etain}
\end{figure}

Applying the jet algorithm,
one obtains either $\gamma+(0+1)$-jet or $\gamma+(1+1)$-jet final states,
with the quark forming a jet on its own in  the latter case. 
As before, these 
are identified as 
$\gamma+(1+1)$-jet events only if the quark jet can be seen inside the 
detector coverage, i.e.\ if
\begin{equation}
E_{T,q} > 2.5~\mbox{GeV}\, , \qquad -2.1<\eta_q<2.1\;.
\end{equation}
Using these cuts, we can define two different isolated photon cross sections:
the isolated $\gamma+(0+1)$-jet cross section, which contains only events 
where no quark jet is observed, and the inclusive isolated 
 $\gamma$ cross section, where no restrictions are applied on the quark jet. 
Note that for the inclusive HERA-frame $k_T$-algorithm, 
these two cross sections would coincide exactly. 
As seen in the previous section, 
results obtained using the inclusive HERA-frame
 $k_T$-algorithm are almost identical to 
results obtained from the exclusive algorithm
for $y_{cut}=0.1$, as applied here. Therefore, there is only very 
little difference between the  isolated $\gamma+(0+1)$-jet cross section
and the isolated  inclusive $\gamma$ cross section for this algorithm.

\begin{figure}[t]
\begin{center}
\epsfig{file=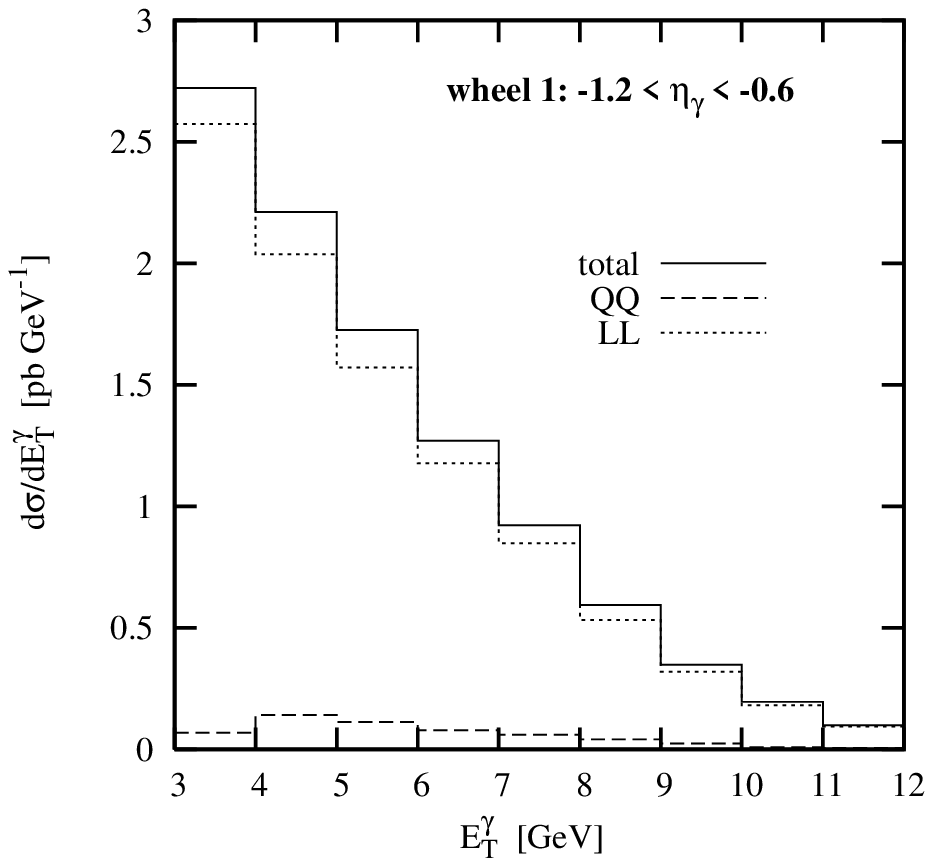,angle=-90,width=8cm} 
\hspace{-1.5cm} \epsfig{file=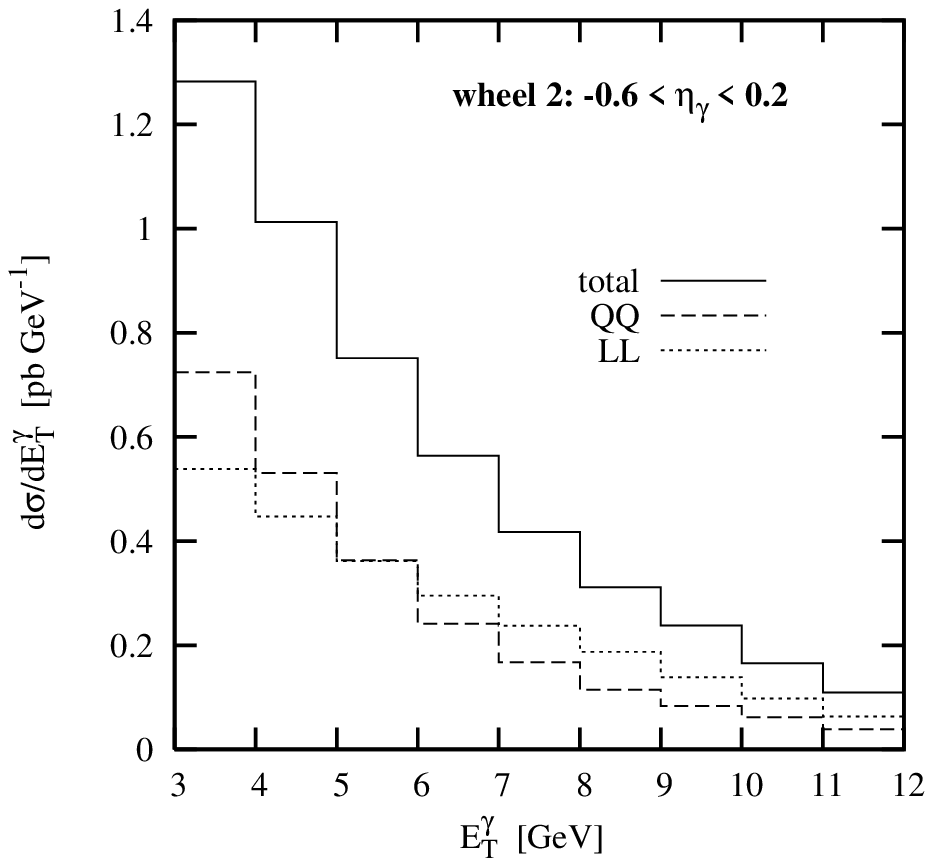,angle=-90,width=8cm}
\\[1cm]
\epsfig{file=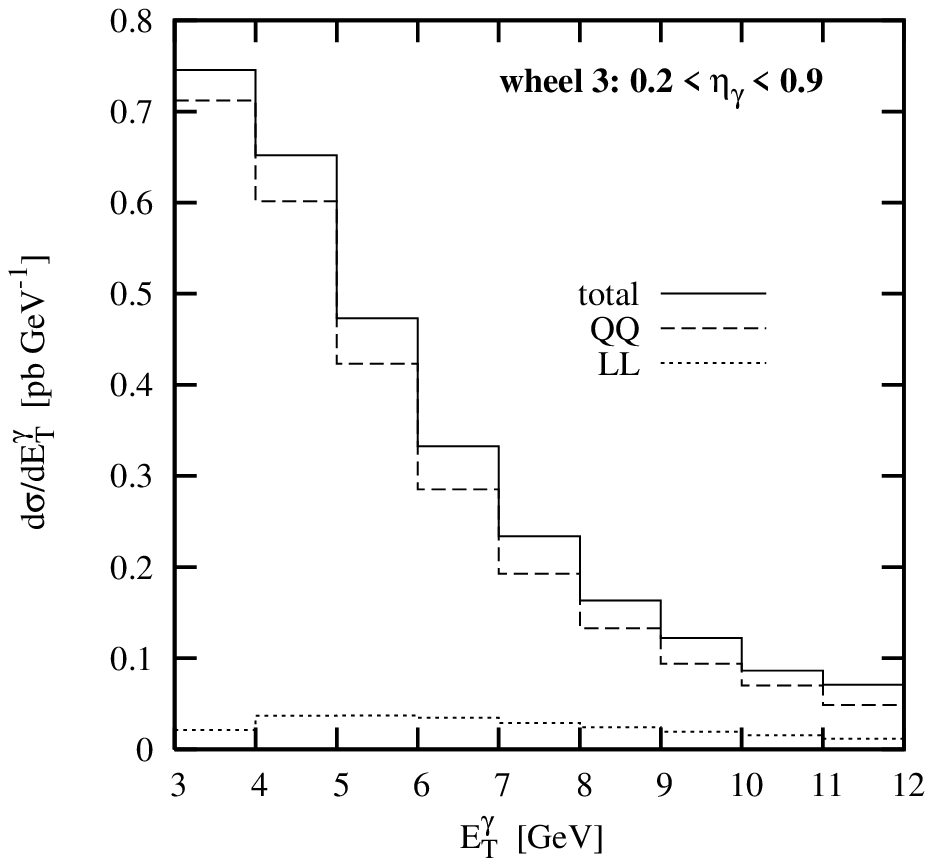,angle=-90,width=8cm} 
\hspace{-1.5cm} \epsfig{file=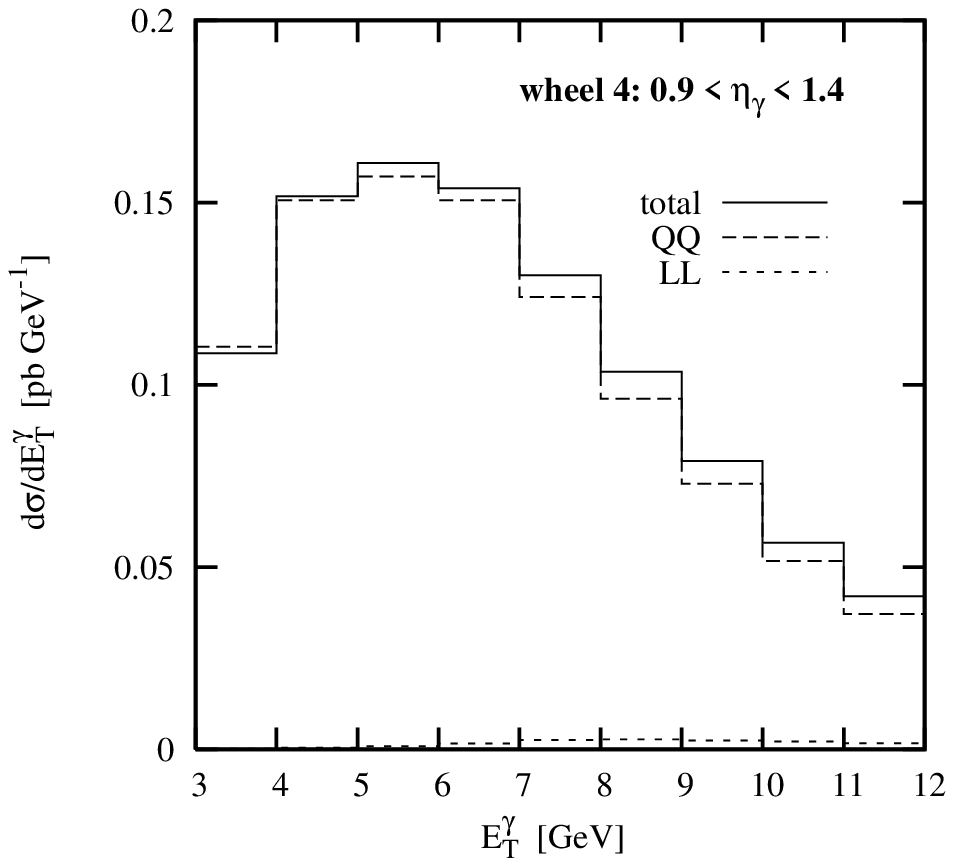,angle=-90,width=8cm}
\\[1cm]
\epsfig{file=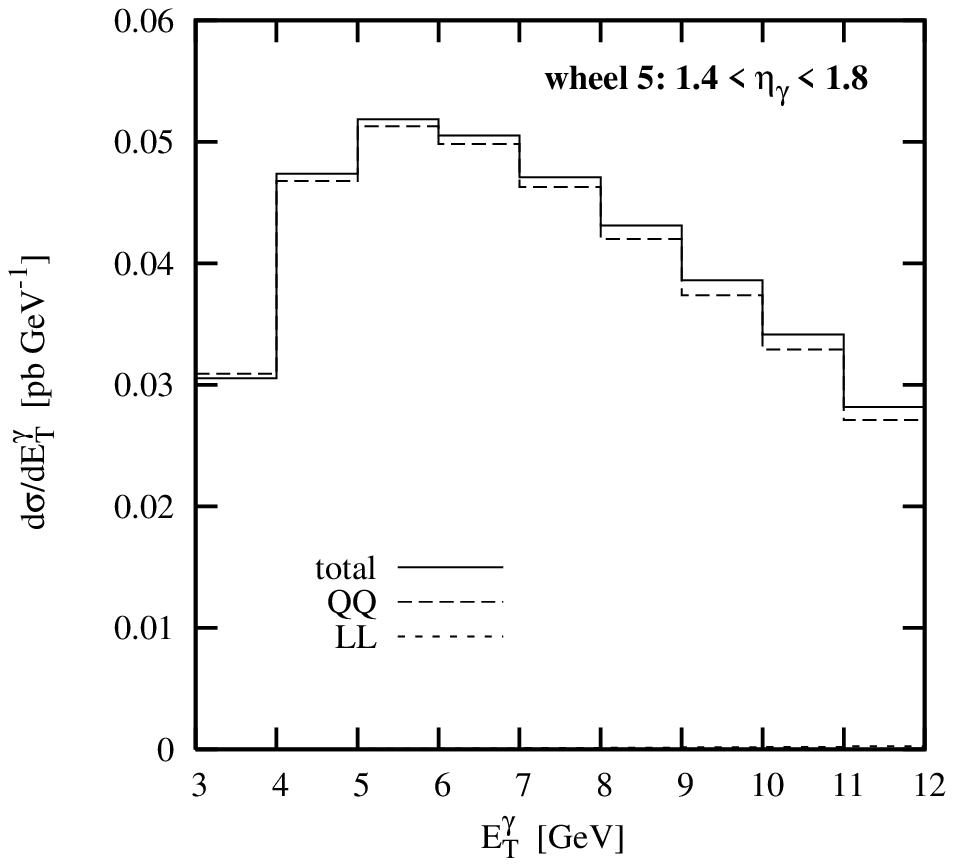,angle=-90,width=8cm} 
\hspace{-1.5cm} \epsfig{file=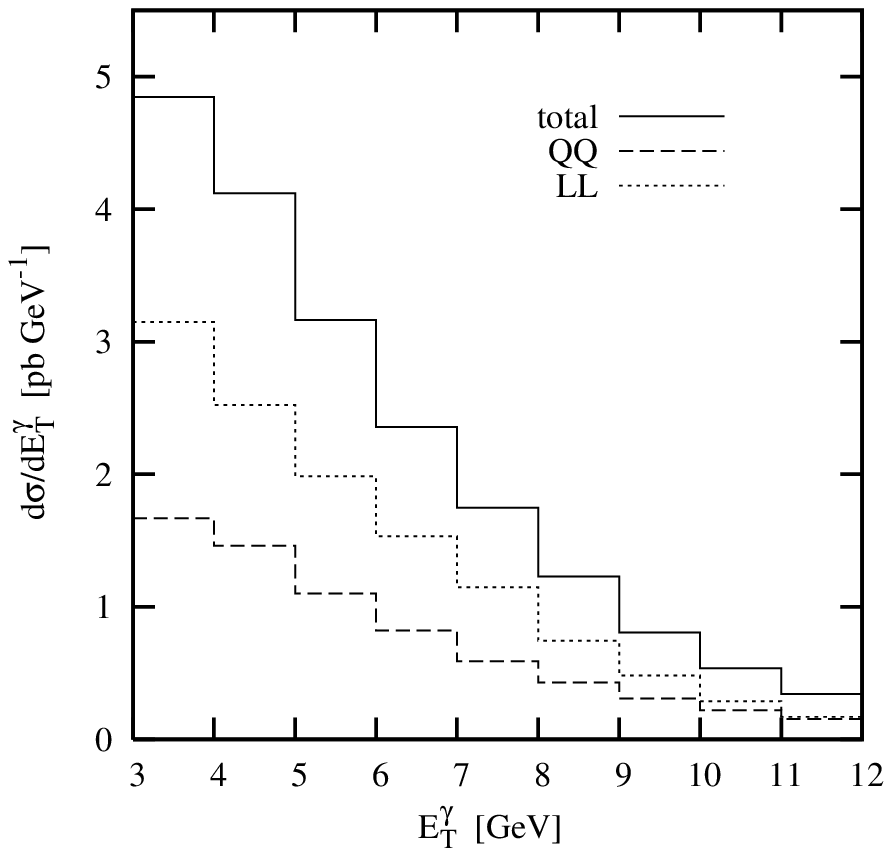,angle=-90,width=8cm}
\end{center}
\caption{Transverse energy 
 distributions of isolated photons in $\gamma+(0+1)$-jet 
events, in different bins 
in $\eta_{\gamma}$. The last plot shows the sum over 
all bins. Isolated photons are 
defined using the exclusive $k_T$-algorithm ($y_{cut} = 0.1$)
in the HERA frame, 
requiring $z>0.9$. $LL$ and 
$QQ$ subprocess contributions are indicated as dashed and dotted lines.}
\label{fig:etin}
\end{figure}
Figures~\ref{fig:etain} and~\ref{fig:etin} 
display the rapidity and 
transverse energy distributions of isolated photons in 
$\gamma+(0+1)$-jet 
events using the exclusive $k_T$-algorithm in the HERA frame. 
$QQ$ and $LL$ contributions to these distributions are indicated separately, 
and the total is obtained by summing $QQ$, $LL$ and $QL$ contributions. 
The total 
$\gamma+(0+1)$-jet 
cross section with this jet algorithm and the above-mentioned cuts
is 19.9~pb.
For the rapidity distributions, we consider three different bins in 
transverse energy, displayed in Figure~\ref{fig:etain}. The rapidity 
distribution of photons 
in $\gamma+(0+1)$-jet production
shows features similar to the rapidity distribution of 
inclusive isolated photons, discussed in~\cite{zeusnew}. 
The distributions resemble each other in all bins 
in $E_{T,\gamma}$, and fall towards increasing 
$\eta_\gamma$. The contributions of the $QQ$ and $LL$ subprocesses are 
of comparable magnitude, but have considerably different shapes in 
$\eta_\gamma$: the $LL$ process is largest in the backward direction
(i.e.\ in the direction of the outgoing electron) and falls rapidly 
towards positive $\eta_{\gamma}$, becoming negligible above $\eta_\gamma
\gapprox 0.5$. The shape of the $LL$ process on one hand, differs very little 
for the different bins. 
The $QQ$ process, on the other hand, is most pronounced at 
mid-rapidity, with a maximum around $\eta_\gamma
\approx 0.5$ for the sum of all  $E_{T,\gamma}$-bins. The position of 
this $\eta_\gamma$-maximum  of the $QQ$ process shifts from lower 
$\eta_\gamma$ in the lowest $E_{T,\gamma}$-bin (where it is around 
$\eta_\gamma\approx 0$)
to higher $\eta_\gamma$ in the highest  $E_{T,\gamma}$-bin 
(maximum around $\eta_\gamma\approx 1$). The three different 
bins are of increasing size, and contribute about equal amounts to the 
total $\eta_\gamma$-distribution. 

\begin{figure}[t]
\begin{center}
\epsfig{file=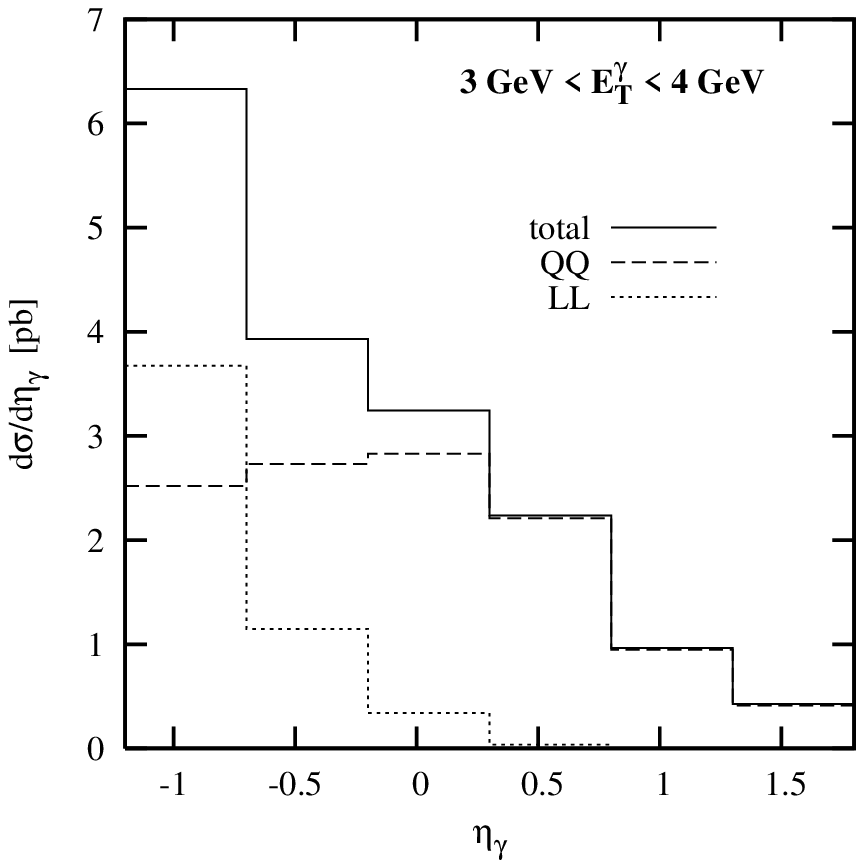,angle=-90,width=8cm} 
\hspace{-1.5cm} \epsfig{file=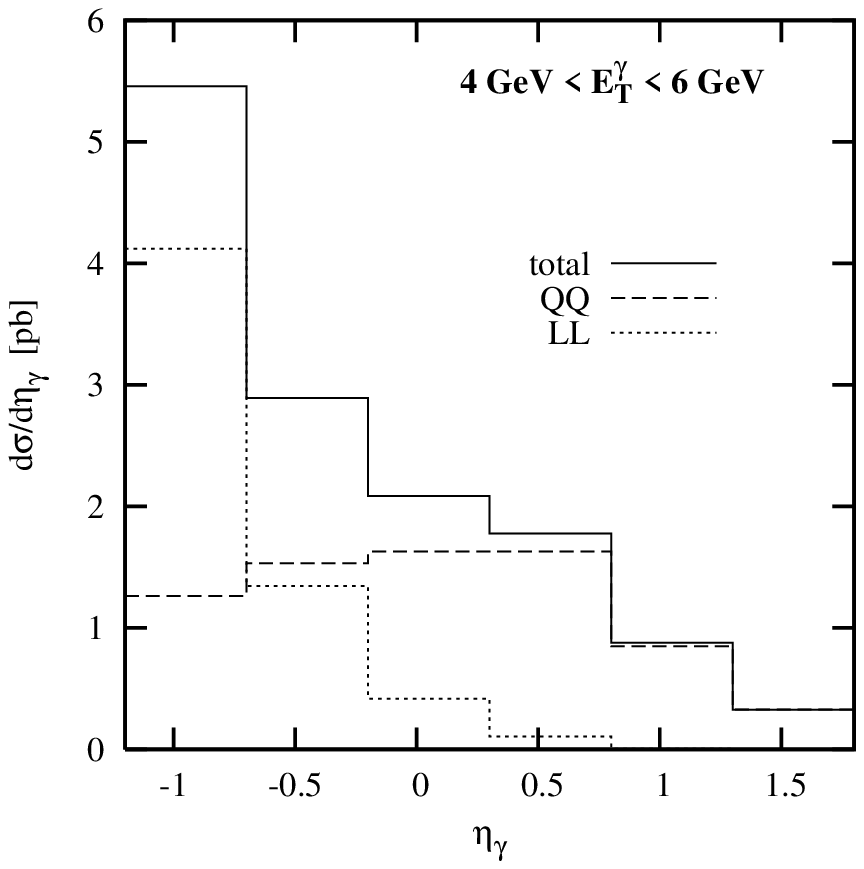,angle=-90,width=8cm}
\\[1cm]
\epsfig{file=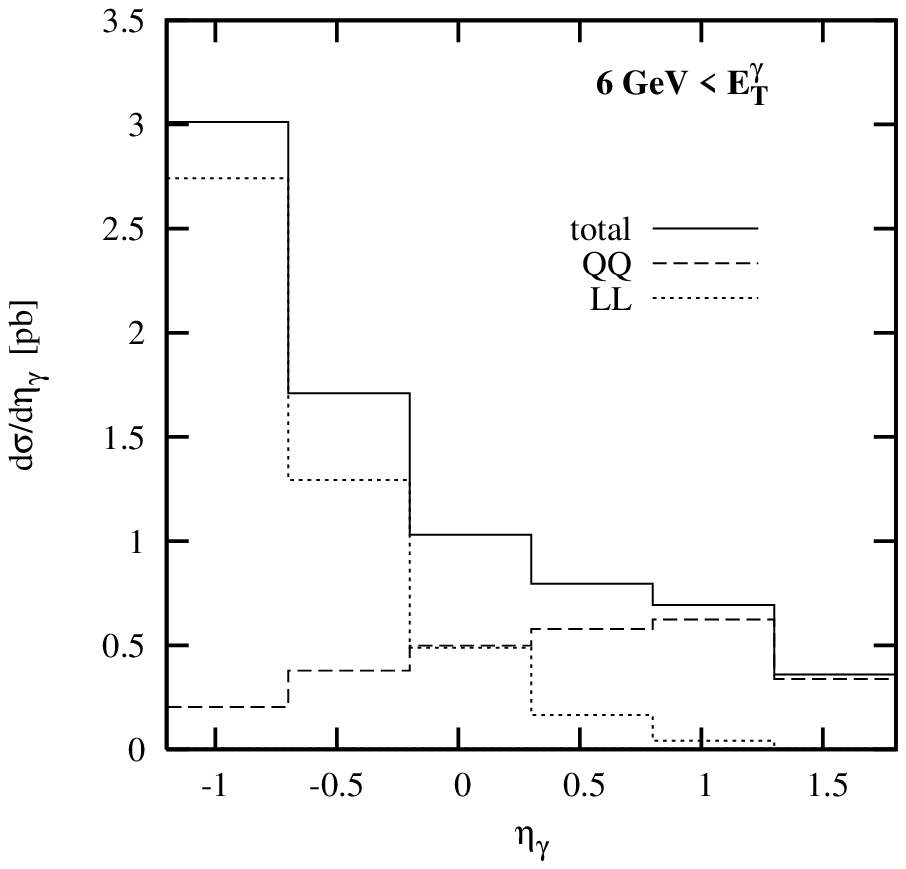,angle=-90,width=8cm} 
\hspace{-1.5cm} \epsfig{file=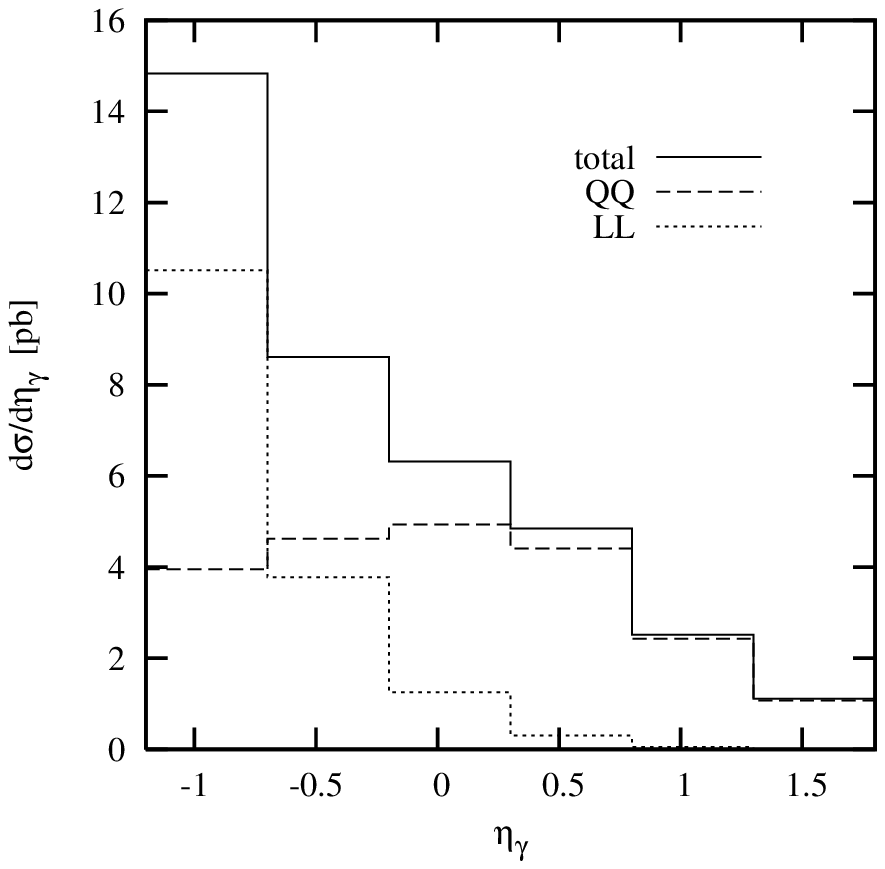,angle=-90,width=8cm}
\end{center}
\caption{Rapidity distributions of isolated photons in $\gamma+(0+1)$-jet 
events, in different bins 
in $E_{T,\gamma}$. The last plot shows the sum over 
all bins. Isolated photons are 
defined here using the hadronic $k_T$-algorithm
($R=1$), requiring $z>0.9$. $LL$ and 
$QQ$ subprocess contributions are indicated as dashed and dotted lines.}
\label{fig:etacone}
\end{figure}
The transverse energy distributions, Figure~\ref{fig:etin},
 are considered in five different 
bins in $\eta_\gamma$, corresponding to five different wheels of the 
electromagnetic calorimeter of the H1 detector~\cite{carsten}.
The numbering of the wheels 
is from the backward towards the forward direction. 
In the first wheel ($-1.2<\eta_\gamma<-0.6$), 
the cross section is completely dominated by the $LL$-process, and 
falls monotonously with $E_{T,\gamma}$. Already in the second wheel
($-0.6<\eta_\gamma<0.2$), $QQ$ and $LL$ processes are of similar magnitude, 
and also of rather similar shape in  $E_{T,\gamma}$. In the third 
wheel ($0.2<\eta_\gamma<0.9$) and beyond, the contribution 
from the $LL$-process is negligible. Like in the first two wheels, 
the $E_{T,\gamma}$-distribution falls monotonously in the 
third wheel. In the fourth ($0.9<\eta_\gamma<1.4$) and fifth 
($1.4<\eta_\gamma<1.8$) wheels, the  $E_{T,\gamma}$-distribution is 
peaked around $E_{T,\gamma}\approx 5.5$~GeV. This feature is a consequence 
of the exclusive HERA-frame $k_T$-algorithm used here: photons produced 
at low transverse energy in the forward region are recombined 
with the proton remnant, and do therefore not contribute to the 
measured cross section. The total transverse energy distribution (summed 
over all wheels in rapidity) is dominated by  the first 
three wheels, and thus receives similar contributions from the $QQ$ and 
$LL$ processes; as always the $QL$ process is of negligible magnitude. 
\begin{figure}[t]
\begin{center}
\epsfig{file=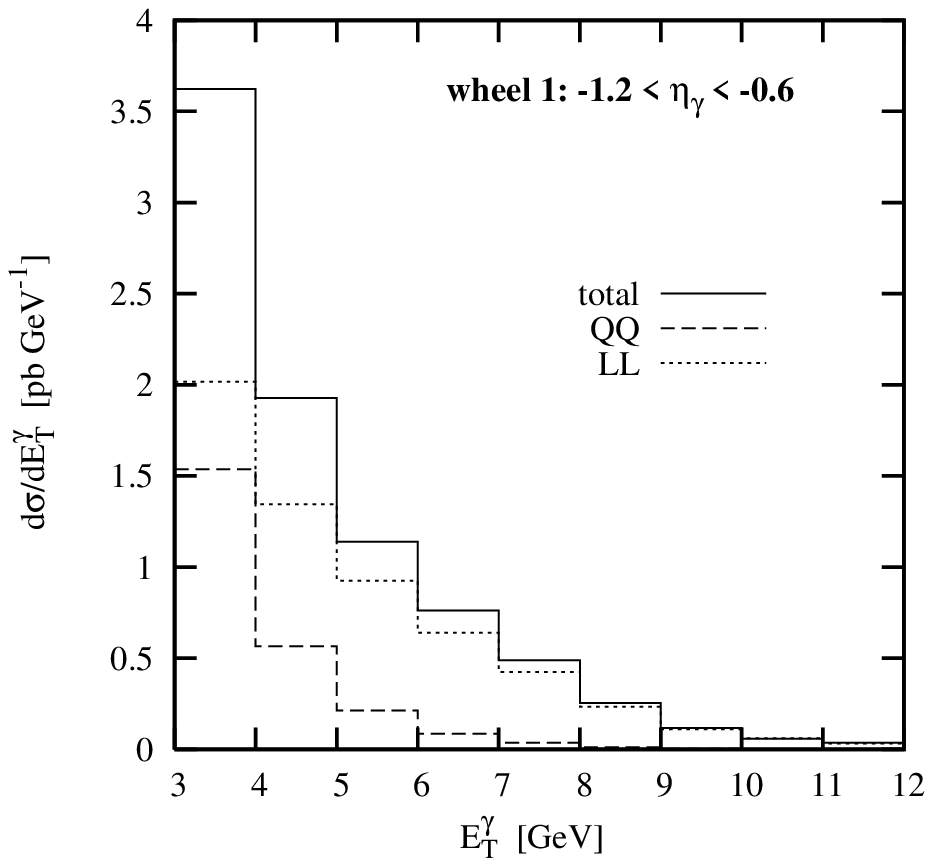,angle=-90,width=8cm} 
\hspace{-1.5cm} \epsfig{file=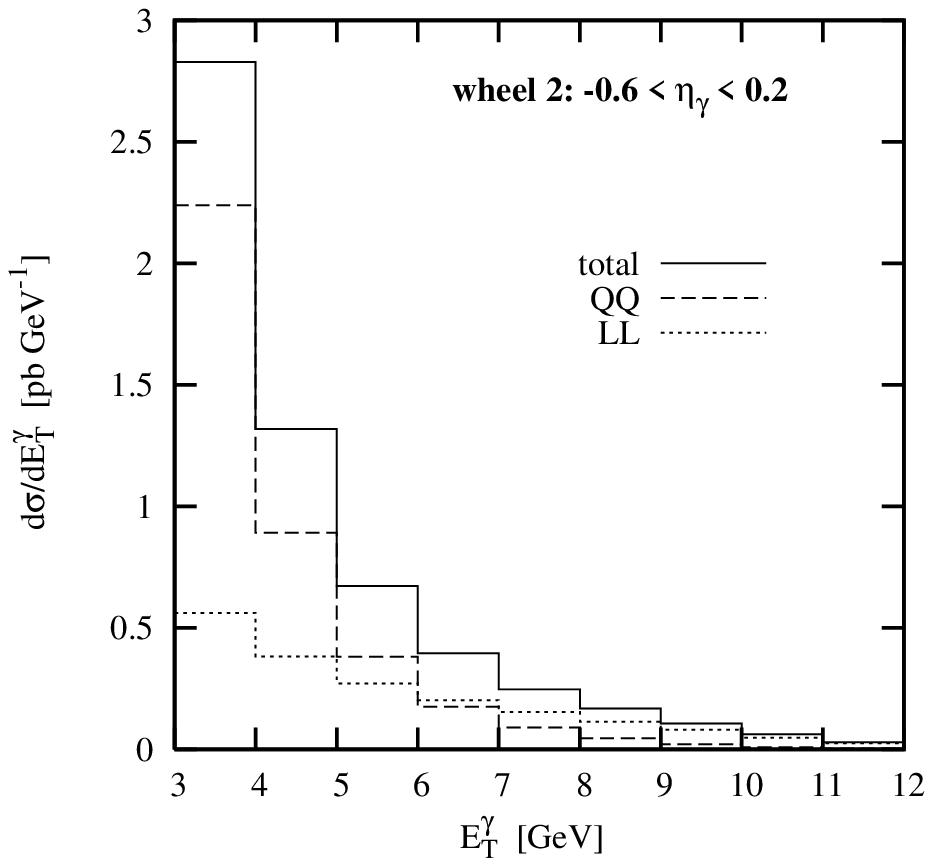,angle=-90,width=8cm}
\\[1cm]
\epsfig{file=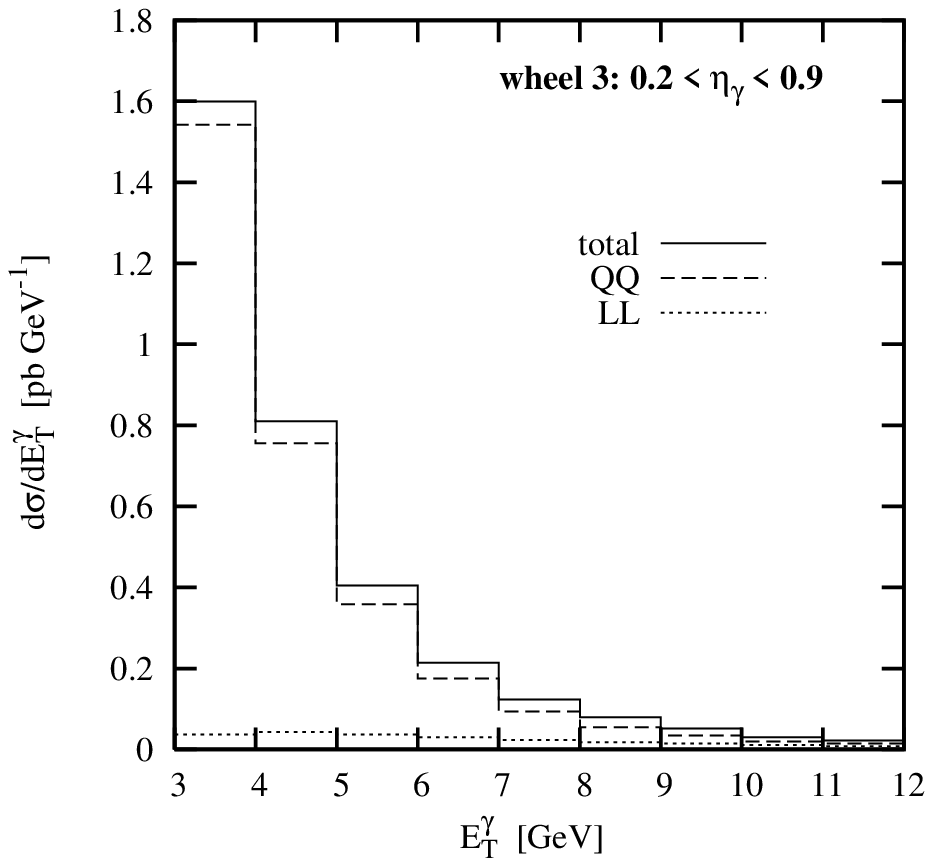,angle=-90,width=8cm} 
\hspace{-1.5cm} \epsfig{file=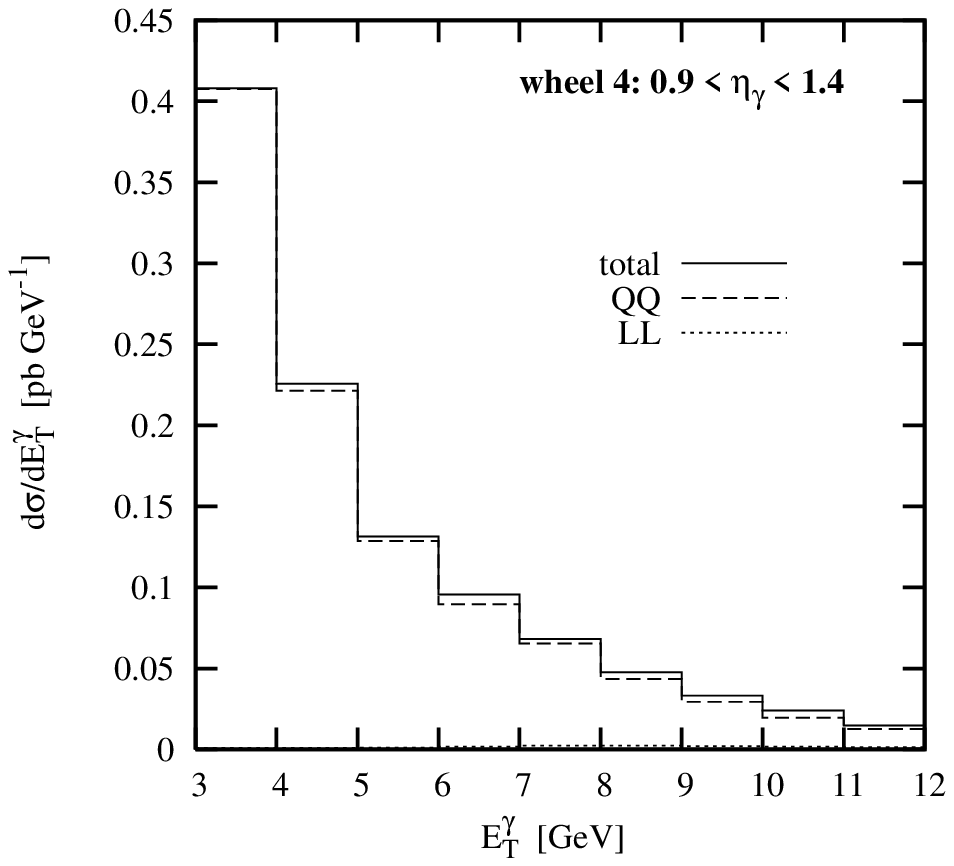,angle=-90,width=8cm}
\\[1cm]
\epsfig{file=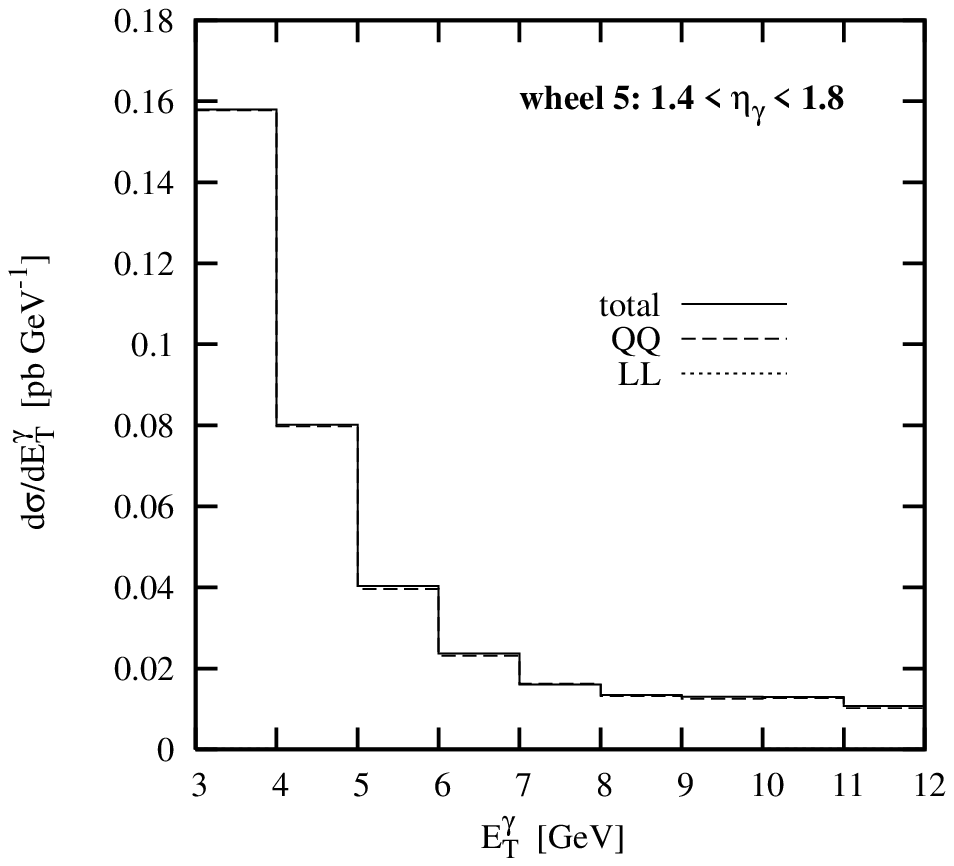,angle=-90,width=8cm} 
\hspace{-1.5cm} \epsfig{file=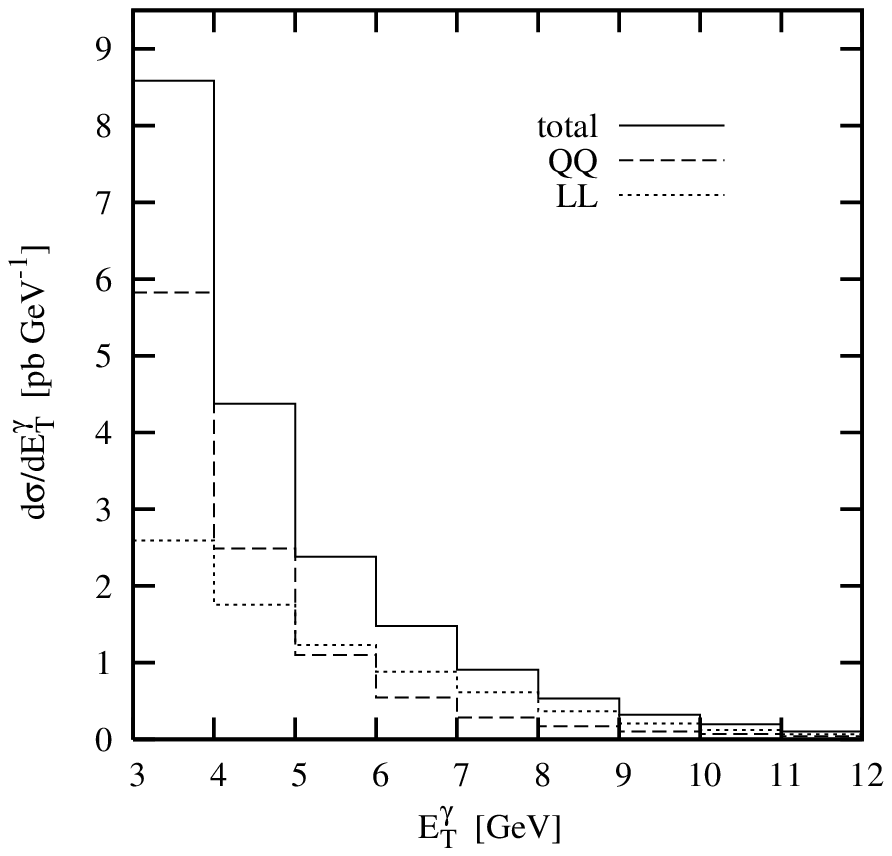,angle=-90,width=8cm}
\end{center}
\caption{Transverse energy 
 distributions of isolated photons in $\gamma+(0+1)$-jet 
events, in different bins 
in $\eta_{\gamma}$. The last plot shows the sum over 
all bins. Isolated photons are 
defined here using the hadronic $k_T$-algorithm ($R=1$), 
requiring $z>0.9$. $LL$ and 
$QQ$ subprocess contributions are indicated as dashed and dotted lines.}
\label{fig:etcone}
\end{figure}

In Figures~\ref{fig:etacone} and~\ref{fig:etcone} 
we show the rapidity and 
transverse energy distributions of isolated photons in 
$\gamma+(0+1)$-jet 
events using the hadronic $k_T$-algorithm ($R=1$). 
As for the HERA-frame exclusive $k_T$-algorithm, 
$QQ$ and $LL$ contributions  are indicated separately, 
and the total is obtained by summing $QQ$, $LL$ and $QL$ contributions. 
We also use the same bins as before. The total  $\gamma+(0+1)$-jet 
cross section with the hadronic $k_T$-algorithm is 19.1~pb, which is 
very similar to the total cross section in the HERA-frame 
exclusive $k_T$-algorithm. Many features of the distributions are 
similar to what we observed above. In the discussion of these figures,
we therefore only focus on differences 
arising from the use of the two different algorithms. 

In the rapidity distributions, Figure~\ref{fig:etacone}, we 
observe that the shape of the $LL$ contribution is similar for both 
jet algorithms, while the $QQ$ contribution looks considerably different.
As opposed to Figure~\ref{fig:etain}, we see that the $QQ$ subprocess 
remains sizable also in the backward rapidity region, especially at low 
$E_{T,\gamma}$. 

The difference between the two jet algorithms is more pronounced in the 
transverse energy distribution, Figure~\ref{fig:etcone}. With 
increasing $E_{T,\gamma}$, this distribution falls more steeply
for the hadronic $k_T$-algorithm than for the HERA-frame exclusive 
$k_T$-algorithm. Also, one observes in the forward region 
(the fourth and fifth wheel) that photons at low transverse energy are 
not disfavoured as in Figure~\ref{fig:etin}, where they 
were combined with the proton remnant in a sizable fraction of the events. 
As a consequence, the total transverse energy distribution falls 
more steeply than for the HERA-frame exclusive 
$k_T$-algorithm.

\begin{figure}[t]
\begin{center}
\epsfig{file=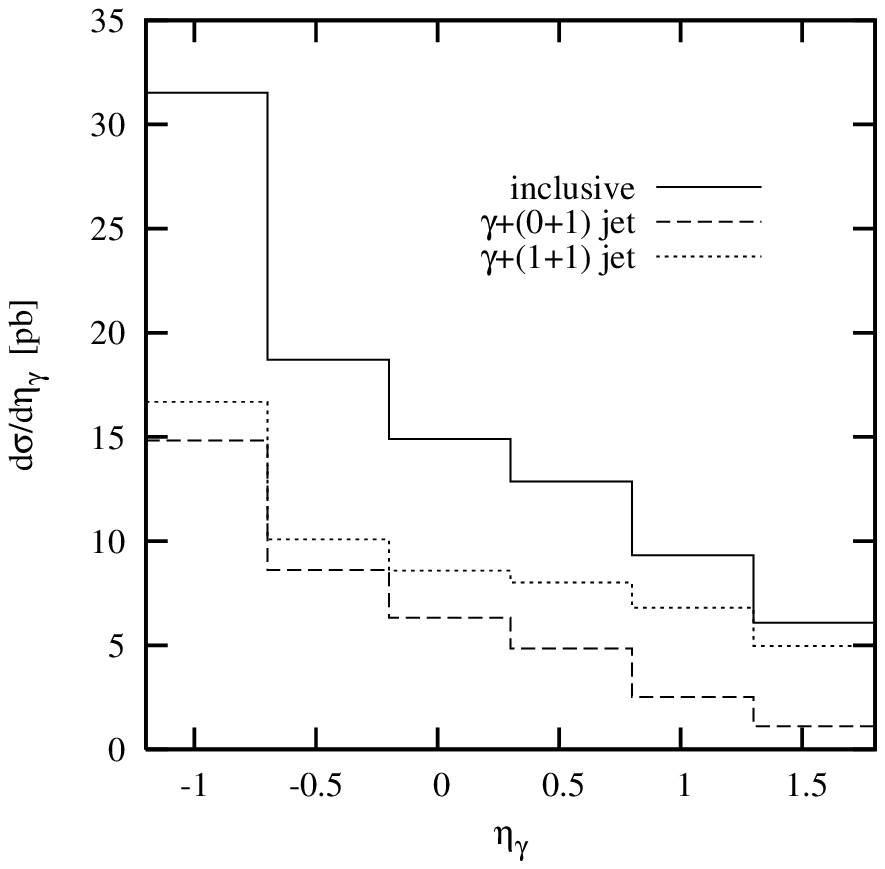,angle=-90,width=8cm} 
\hspace{-1.5cm} \epsfig{file=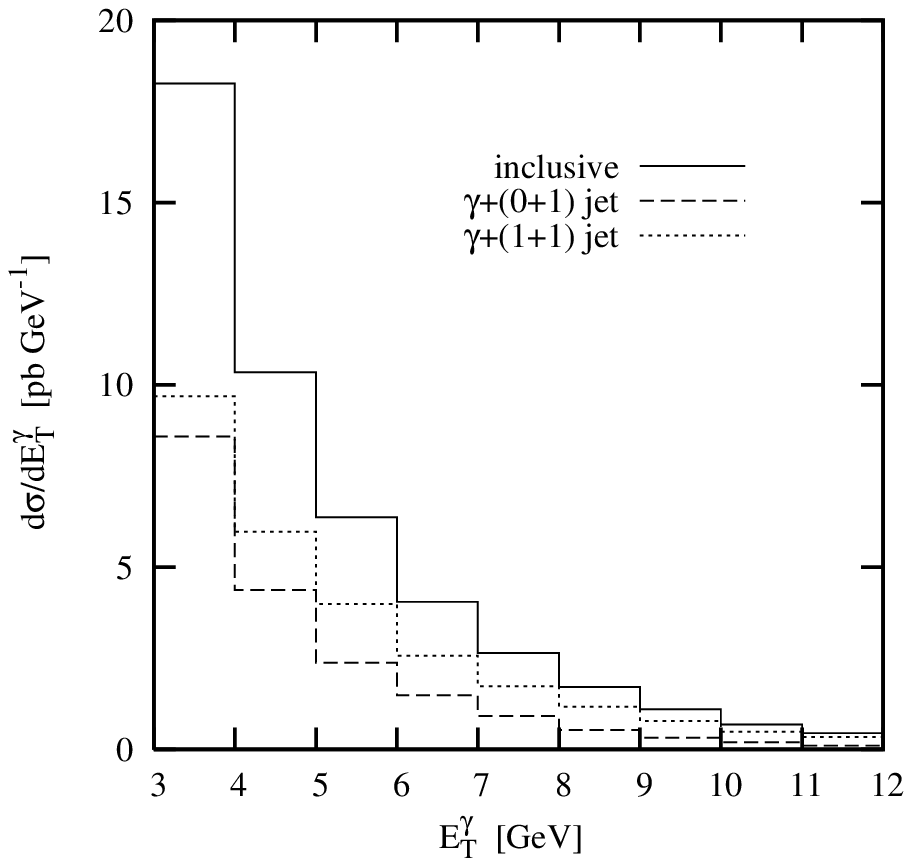,angle=-90,width=8cm}
\end{center}
\caption{Comparison of isolated 
$\gamma+(0+1)$-jet, $\gamma+(1+1)$-jet
and inclusive isolated $\gamma$ cross sections using the hadronic 
$k_T$-algorithm ($R=1$), requiring $z>0.9$.}
\label{fig:incljet}
\end{figure}
As explained above, the exclusive $k_T$-algorithm in the HERA frame 
almost 
always yields $\gamma+(0+1)$-jet final states, such that in this algorithm 
the isolated $\gamma+(0+1)$-jet cross section nearly coincides with the 
inclusive isolated $\gamma$ cross section. In contrast, 
application of the hadronic 
$k_T$-algorithm results in $\gamma+(0+1)$-jet  and 
$\gamma+(1+1)$-jet final states. At the leading order in perturbation theory 
used here, the inclusive  isolated $\gamma$ cross section in 
this algorithm is the sum of the $\gamma+(0+1)$-jet and 
$\gamma+(1+1)$-jet cross sections. 
The inclusive isolated $\gamma$ cross section 
and its decomposition into $\gamma+(0+1)$-jet  and 
$\gamma+(1+1)$-jet final states is shown in Figure~\ref{fig:incljet}. For the 
integrated cross sections, we obtain 19.1~pb for $\gamma+(0+1)$-jet,
27.6~pb for $\gamma+(1+1)$-jet and thus 46.7~pb for the inclusive 
cross section. This cross section is thus considerably larger than the 
inclusive isolated $\gamma$ cross section obtained with the 
exclusive HERA-frame $k_T$-algorithm. As already discussed in 
section~\ref{sec:frag}, the latter algorithm is more likely to
cluster photon and quark together into
a single jet. Consequently, many final state configurations that were
identified as photon jets with 
$z=1$ by the hadronic $k_T$-algorithm yield  photon jets with $z<1$ 
with the exclusive HERA-frame $k_T$-algorithm. If these photon jets have 
$z<0.9$, they do no longer contribute to the isolated photon cross section.

We observe that the  $\gamma+(1+1)$-jet distributions fall less steeply 
than the $\gamma+(0+1)$-jet distributions, both in rapidity and in transverse 
energy. This feature can be understood from the fact that at large forward 
rapidity or at large transverse energy, it is kinematically preferred that the 
transverse energy of the photon is balanced by both the electron and the 
hard jet. Also, the $\gamma+(1+1)$-jet cross section exceeds the 
$\gamma+(0+1)$-jet everywhere in phase space. This might appear 
counter-intuitive at first sight, but may be understood from the fact that 
both cross sections start at the same order in perturbation theory,
namely ${\cal O}(\alpha^3)$. The admixture of  
$\gamma+(0+1)$-jet and $\gamma+(1+1)$-jet events in the inclusive 
sample is highly dependent on the cuts applied to the quark jet, especially 
on its transverse energy cut, which is chosen here to be even lower than 
the cut on the transverse energy of the photon. 

As a final point, we note that the measurement of the ZEUS 
collaboration~\cite{zeusnew}, based on a cone-based photon isolation,
 yielded an inclusive isolated photon production cross section 
considerably larger than the 
 $\gamma+(1+1)$-jet cross section. This behaviour is due to the 
more restrictive cuts on the hadronic jet applied by ZEUS
to select  $\gamma+(1+1)$-jet final states. 
 Both ZEUS measurements are in good agreement 
with the theoretical approach advocated here: we compared the inclusive 
isolated photon cross section with the ZEUS measurement in~\cite{zeusnew},
and ZEUS compared~\cite{zeus} their measurement of the 
 $\gamma+(1+1)$-jet cross section to an earlier NLO calculation~\cite{herajet},
based on the same approach which we used here at leading order. 

\section{Conclusions and Outlook}
\setcounter{equation}{0}
\label{sec:conc}

In this paper, we studied the production of final state photons in 
deep inelastic scattering at leading order in perturbation 
theory, ${\cal O}(\alpha^3)$. Already at this 
leading order, the corresponding 
parton-level cross section contains a collinear quark-photon divergence, 
which is absorbed into the quark-to-photon fragmentation function.
Our calculation of final state photon production contains therefore 
both contributions from hard parton-level  photon radiation and 
from photon fragmentation. 

Besides a perturbatively generated component, the quark-to-photon 
fragmentation function contains a genuinely non-perturbative component, 
which forms the boundary condition to its perturbative evolution equation. 
Experimental measurements of this photon fragmentation function 
were made up to now only in electron-positron annihilation at 
LEP~\cite{aleph,opal}.

In the democratic clustering procedure~\cite{glover-morgan} for 
photon cross sections, the photon candidate is clustered by the jet algorithm 
like any hadron in the event. As a result, one of the final state 
jets contains a highly energetic photon, and is called photon jet, 
abbreviated by $\gamma$. Using this procedure, we studied the  
 $\gamma+(0+1)$-jet production cross section
in deep inelastic scattering at HERA, and demonstrated that the energy 
distribution of photons inside the photon jet in these events is 
highly sensitive on the quark-to-photon fragmentation function, and 
can be used to discriminate different available parametrisations of it. 
We could show that such a measurement is best carried out using a
particular variant of the $k_T$-algorithm, which enhances the importance 
of fragmentation contributions relative to the hard radiation. 

Isolated photons are usually defined at high energy experiments by
allowing them to be accompanied by some amount of hadronic energy, since 
a perfectly isolated photon is not infrared safe in perturbation theory. 
The democratic clustering procedure allows a natural definition of isolated 
photons by identifying the photon jet as isolated photon if the 
fraction of its energy carried by the photon candidate exceeds some 
value defined by the experimental environment. At HERA, photons are 
called isolated if they carry more than 90\% of the transverse energy of 
the photon jet. 

Using this definition, we studied isolated photon cross sections for 
 $\gamma+(0+1)$-jet, $\gamma+(1+1)$-jet and inclusive $\gamma$ final states
for different jet algorithms. We found that particular features of 
the parton-level processes and of the jet algorithm can be related to
aspects of the rapidity and transverse energy 
distributions of the photons. 

As in our previous study of isolated inclusive photon production 
in deep inelastic scattering~\cite{zeusnew}, based on a cone-based 
isolated criterion used in the corresponding experimental 
measurement~\cite{zeus}, we found that photon radiation off the lepton and 
off the quark are of comparable  importance, although either of them 
dominates in a different region in photon rapidity. This has important 
implications for the use of inclusive photon cross sections to measure 
the photon distribution in the proton~\cite{pisano,mrst}, as needed for 
electroweak corrections to hadron collider observables~\cite{phodist}. 
In particular, 
it invalidates the assumption~\cite{mrst} that the bulk of 
the isolated inclusive 
photon cross section in DIS arises only 
from photon radiation off 
the lepton, as already pointed out in~\cite{saxon,zeusnew}. If possible at 
all, an extraction of the photon distribution in the proton would have to
be restricted to kinematical regions where radiation off the lepton is 
indeed dominant. 

NLO corrections, ${\cal O}(\alpha^3\alpha_s)$, are known to the 
$\gamma+(1+1)$-jet cross section in deep inelastic scattering~\cite{herajet}
for some time already; this calculation was found in good agreement 
with experimental data recently~\cite{zeus}. The derivation of NLO 
corrections to the $\gamma+(0+1)$-jet cross section and the 
inclusive photon cross section in deep inelastic scattering is 
however considerably 
more involved. Owing to the appearance of the collinear quark-photon 
singularity in these observables already at leading order, an NLO calculation 
will encounter double unresolved partonic configurations, which are otherwise
expected only at NNLO. In this sense, such a calculation would 
have similar features as the calculation of the NLO corrections 
to the $\gamma+1$-jet rate at LEP~\cite{agg}, 
where first developments towards double unresolved real radiation were made.

\section*{Acknowledgements} 
We would like to thank Katharina M\"uller, Carsten Schmitz, 
 Ulrich Straumann and David Saxon
for many useful and clarifying discussions. 
This work was supported by the Swiss National Science Foundation 
(SNF) under contract PMPD2-106101.

\end{document}